\documentclass[final, 3p, times]{elsarticle}
\usepackage{setspace}

\usepackage{mathpazo}

\usepackage[]{fontenc}
\usepackage[latin9]{inputenc}

\usepackage[colorlinks]{hyperref}
\AtBeginDocument{%
\hypersetup{citecolor=magenta,breaklinks,urlcolor=blue,linkcolor=red}
}%

\usepackage{graphics}
\usepackage{graphicx}
\usepackage{epsfig}
\usepackage{epstopdf}
\usepackage{float}
\usepackage{amssymb}
\usepackage{pifont}
\usepackage{amsfonts}
\usepackage{amsmath}
\numberwithin{equation}{section}
\usepackage{lineno}
\modulolinenumbers[10]

\usepackage{bbding}
\usepackage{multirow}
\usepackage{verbatim} % This package has the option to comment large sections \begin{comment} bla bla bla \end{comment}

\newcommand{\be}{\begin{equation}}
\newcommand{\ee}{\end{equation}}
\newcommand{\DD}{\mathcal{D}}
\newcommand{\PP}{\mathcal{P}}
\newcommand{\MM}{\mathcal{M}}
\newcommand{\bfb}{\boldsymbol{\beta}}
\newcommand{\bfg}{\boldsymbol{\gamma}}
\journal{Astronomy and Computing}

\begin{document}

\begin{frontmatter}



%\title{Comparison of Bayesian Model and Variable Selection without Evidence for the Gaussian Linear Model}

\title{Bayes Factors via Savage-Dickey Supermodels}


\author[uct,aims,saao]{A.~Mootoovaloo\corref{cor1}\fnref{fn1}}%
%\ead{arrykrish@gmail.com}
\author[uct,aims,saao]{Bruce A.~Bassett\corref{cor2}\fnref{fn2}}%
%\ead{bruce.a.bassett@gmail.com}
\author[aims,ug]{M.~Kunz\fnref{fn3}}%
%\ead{martin.kunz@unige.ch}

\cortext[cor1]{Corresponding author}
\cortext[cor2]{Principle Corresponding author}
\fntext[fn1]{\href{mailto:arrykrish@gmail.com}{arrykrish@gmail.com} (A.Mootoovaloo)}
\fntext[fn2]{\href{mailto:bruce.a.bassett@gmail.com}{bruce.a.bassett@gmail.com} (B.A.Bassett)}
\fntext[fn3]{\href{mailto:martin.kunz@unige.ch}{martin.kunz@unige.ch} (M.Kunz)}

\address[uct]{Department of Mathematics and Applied Mathematics, University of Cape Town,
Rondebosch, Cape Town, 7700, South Africa}
\address[aims]{African Institute for Mathematical Sciences, 6 Melrose Road, Muizenberg, 7945, South Africa}
\address[saao]{South African Astronomical Observatory, Observatory Road, Observatory, Cape Town, 7935,
South Africa}
\address[ug]{D\'{e}partement de Physique Th\'{e}orique and Center for Astroparticle Physics, Universit\'{e} de
Gen\`{e}ve, Quai E. Ansermet 24, CH-1211 Gen\`{e}ve 4, Switzerland}

\begin{abstract}
We outline a new method to compute the Bayes Factor for model selection which bypasses the Bayesian Evidence. Our method combines multiple models into a single, nested, Supermodel using one or more hyperparameters. Since the models are now nested the Bayes Factors between the models can be efficiently computed using the Savage-Dickey Density Ratio (SDDR). In this way model selection becomes a problem of parameter estimation.  We consider two ways of constructing the supermodel in detail: one based on combined models, and a second based on combined likelihoods. We report on these two approaches for a Gaussian linear model for which the Bayesian evidence can be calculated analytically and a toy nonlinear problem. Unlike the combined model approach, where a standard Monte Carlo Markov Chain (MCMC) struggles, the combined-likelihood approach fares much better in providing a reliable estimate of the log-Bayes Factor. This scheme potentially opens the way to computationally efficient ways to compute Bayes Factors in high dimensions that exploit the good scaling properties of MCMC, as compared to methods such as nested sampling that fail for high dimensions.  
\end{abstract}


\end{frontmatter}

%\linenumbers

\section{Introduction}

One of the key questions underlying science is that of model selection: how do we select between competing theories which purport to explain observed data? The great paradigm shifts in science fall squarely into this domain. In the context of astronomy - as with most areas of science - the next two decades will see a massive increase in data volume through large surveys such as the Square Kilometre Array (SKA) \citep{hollitt2016overview} and LSST \citep{becla2006designing}. 
Robust statistical analysis to perform model selection at scale will be a critical factor in the success of such future surveys. 

The basic problem of model selection is easy to state. As one considers models with more and more free parameters, one must expect that such models will fit any dataset better and better, irrespective of whether they have anything to do with reality. This problem of overfitting has led to many proposed methods to deal with this kind of situation: that is, finding a way to suitably penalise extra parameters. One method is LASSO (Least Absolute Shrinkage and Selection Operator)  \citep{hastie2005elements}. Other methods such as Akaike Information Criterion (AIC) \citep{akaike1974new} and Bayesian Information Criteria (BIC) \citep{schwarz1978estimating} penalise the best fit likelihood based on the number of free parameters \citep{gelman2014understanding}.

From a Bayesian point of view, model selection is not viewed as a question to be answered looking only at a single point in the parameter spaces, e.g. the point of maximum likelihood of the models in question, but rather should also depend on the full posterior distribution over the parameters. Hence selection is performed by choosing the model with the maximum model  probability $\PP(\MM|\DD)$, derived from the Bayesian Evidence (or marginal likelihood) $\PP(\DD|\MM)$. This automatically expresses Occam's razor, thus penalising extra parameters
which are not warranted by the data. 

Here and throughout this paper we will use $\DD$ to denote data and $\MM$ for a model. Given two competing models, one
would typically compute the Bayesian Evidence for each model and hence
the Bayes Factor, which is the ratio of the evidences.  There are a number of issues  with the Bayesian evidence. It is very sensitive to priors and, of key interest to us, since it involves integrals over the full parameter spaces of each model, is hard to compute efficiently. Techniques such as nested sampling (\citep{skilling2004nested}) scale exponentially with the number of parameters and cannot be used for high-dimensionality problems.  

However, if one model is nested within the other (i.e. all the parameters of one model are shared by another), we can use the Savage-Dickey Density Ratio (SDDR) (\citet{dickey1971weighted} and \citet{verdinelli1995computing})
to directly calculate the Bayes Factor. As an example, consider the
case where the parameters in model $\MM_{1}$ are $\phi$
and $\theta$ while the parameter in model $\MM_{2}$ is $\theta$.
Then, $\MM_{2}$ is nested in $\MM_{1}$ at some value of $\phi$ which we can take to be $\phi=0$.
The Bayes Factor is then given directly by 
\begin{equation}
\textrm{B}_{21}=\left.\dfrac{\mathcal{P}\left(\phi\left|\DD,\,\MM_{1}\right.\right)}{\mathcal{P}\left(\phi\left|\MM_{1}\right.\right)}\right|_{\phi=0}
\end{equation}

where $\mathcal{P}\left(\phi\left|\DD,\,\MM_{1}\right.\right)$
is simply the normalised posterior probability distribution of $\phi$ in the
extended model, that is:
\[
\mathcal{P}\left(\phi\left|\DD,\,\MM_{1}\right.\right)=\int\mathcal{P}\left(\theta,\,\phi\left|\DD,\,\MM_{1}\right.\right)\,d\theta
\]

The core of this paper is the idea that it is possible to embed any two models into a {\em Supermodel} such that each model is nested within the supermodel. Related ideas can be found in \citep{hee2016bayesian, hlozek2012photometric, kamary2014testing}. 

In the next sections, we shall illustrate this in detail. The paper is organised as follows: in \S\ref{sec:methods}, we describe our idea in the general context. In \S\ref{sec:application_linear_model} and \S\ref{sec:application_non_linear}, we test our approach using both the linear and non-linear models while in \S\ref{sec:reparam_exploit}, we also consider one example of reparameterization of $\alpha$, the hyperparameter with respect to which the models are nested. We conclude in \S\ref{sec:conclusion}.

\section{Our Methods}
\label{sec:methods}

In this section, we discuss the methods that we shall use to calculate
the Bayes Factor. The key driver of our interest in these methods is the desire for techniques that do not scale exponentially with the complexity of the models, as occurs for nested sampling \citep{feroz2013importance}. 

Monte Carlo Markov Chain (MCMC) itself is useful as a method exactly because it does not scale exponentially with increasing numbers of parameters, and hence our goal is to use MCMC-based methods to compute the Bayes factor. Of course, as with any such method, convergence needs to be achieved and there is some evidence that our supermodel methods do make the posterior harder to sample from with chains that have larger correlation lengths. Nevertheless, since our methods are fundamentally based on MCMC we argue they will still have better scaling properties than nested sampling. Let us now discuss and illustrate the methods in detail.   

The key idea is to embed the models under consideration within a single Supermodel and then use the SDDR to evaluate the Bayes Factor. The embedding of the models can be done in at least two ways. One approach is to embed at the level of the models, another is at the level of the likelihoods. We call these two approaches the {\em Combined Model} and {\em Combined Likelihood} methods. We test both approaches, finding that the Combined Likelihood approach has significant performance advantages. 

\subsection{General Approach\label{sec:general}}

In order to use the SDDR for model selection or comparison even in the case
of non-nested models, we introduce a
hyperparameter, which we denote $\alpha$, that takes on particular values
for the two models that we want to compare (e.g.\ 0 and 1). 
So if we want to compare model $\MM_1$ with model $\MM_2$, we
construct a Supermodel that contains the
sets of parameters $\bfb$ and $\bfg$ of the models $\MM_1$ and $\MM_2$ respectively,
as well as a `nesting parameter' $\alpha$, and that recovers each of the models at $\alpha = 0, 1$ respectively. Namely it satisfies:
\be
\PP_S(\DD | \bfb,\bfg, \alpha = 0) = \PP(\DD|\bfb, \MM_1) \, , \quad
\PP_S(\DD | \bfb,\bfg, \alpha = 1) = \PP(\DD|\bfg, \MM_2) \, . \label{eq:general_condition}
\ee
where $\PP_S(\DD | \bfb,\bfg, \alpha)$ is the supermodel posterior. 
There are a potentially infinite number of supermodels that can achieve this. In 
this paper we restrict ourselves to study of the simplest, linear, implementations, (see eq. 
 (\ref{comblike}), (\ref{combmod})). 
 
The priors for $\MM_\alpha$ additionally need to be chosen so that they
correspond to the desired priors for $\MM_1$ and $\MM_2$ when $\alpha=0$ and
$1$ respectively. One way to do this is to have separable priors under each model such that the parameters corresponding to a specific model are integrated out relatively easily. Alternatively, one can even combine the models via both the likelihoods and the priors.
In this way the models $\MM_1$ and $\MM_2$ are effectively nested inside
the model $\MM_\alpha$ for the purpose of the likelihoods, and we can use the SDDR
to compute the Bayes factor between these two models,
\be
B_{12} = \frac{B_{1\alpha}}{B_{2\alpha}}
= \frac{\PP_S(\alpha=0|\DD)}{\PP_S(\alpha=0)} 
\frac{\PP_S(\alpha=1)}{\PP_S(\alpha=1|\DD)}  \, .
\label{eq:bayesfac}
\ee

\subsubsection{Transformations of $\alpha$ and model averaged posteriors\label{sec:reparam}}

In addition, given a supermodel one can also use any transformation, $\alpha \rightarrow f(\alpha)$
as long as $f(\alpha)$ can take the values $0$ and $1$ within the domain of definition of $\alpha$, so that Eq. (\ref{eq:general_condition}) holds.
In actual applications these limits do not even need to be strictly verified; for example using
$\alpha \rightarrow f(\alpha) = e^{\alpha}$ for $\alpha\in[-\Lambda,0]$ is good enough
for a large enough $\Lambda$, under the (usually true) assumption that the likelihood
$P_S(\DD | \alpha, \bfb,\bfg)$ tends in a continuous way to the limit $P(\DD|\bfb, \MM_1)$
as $f(\alpha)\rightarrow 0$. See Section \ref{sec:reparam_exploit} for a detailed investigation.

In the above we have tacitly assumed that $\alpha$ is a continuous parameter. This is however
not necessary, $\alpha$ can also be an index variable that takes discrete values. This case can be seen
as the limit of a continuous $\alpha$ that has the form of a step function (or a hyperbolic tangent function
with a sharper and sharper transition). In the discrete case, there is not even a need to
explicitly construct a supermodel, as we are always only in one of the simpler models $\MM_1$
or $\MM_2$; see e.g. \citep{hee2016bayesian}.

This limit is also interesting for another reason. It may be that we are not really interested in precise
model probabilities, but rather we want to infer parameter constraints in situations where the model
is uncertain. An example could be image reconstruction, e.g. in astronomy, with an unknown number
of point sources. In this situation our object of interest is the model-averaged posterior for a parameter $\theta$,
\be
\PP(\theta|\DD) = \frac{\sum_j \PP(\theta|\DD,\MM_j) \PP(\MM_j|\DD)}{\sum_j \PP(\MM_j|\DD)} \, .
\ee
From Equation (\ref{eq:bayesfac}) we can see that the Bayes factor $B_{12}$ between two models
is given by the probability to find $\alpha=0$ or $\alpha=1$ if both have equal prior probabilities.
This means that the case where $\alpha$ is indicator variable will directly give us model-averaged
posteriors if we marginalize over all parameters except $\theta$ (but including $\alpha$), without
having to compute $B_{12}$ explicitly.

\subsubsection{More than two models}

There are many different possibilities to deal with more than two models. They could be nested at different
values of a single parameter $\alpha$. Alternatively we can introduce a separate parameter $\alpha_i$ for each
model together with the global constraint $\sum_i \alpha_i = 1$. In this way the space of the $\alpha_i$ forms a
simplex which can be parameterised, for example, with barycentric coordinates and on which an MCMC can move.
The second approach has the advantage that each model can be reached from any point in the simplex without having
to pass through potentially prohibitively bad regions in the global parameter space. On the other hand, we need
to introduce nearly as many new parameters as we have models. In general it is unclear which of these two approaches is superior and leave the study of multiple models to future work. 

\subsubsection{Using the Same Parameters vs Different Parameters}

One of the fundamental choices when using the supermodel approach is how to deal with common parameters to the two models. There are again two options: to explicitly share the common parameters or to decouple the models by replicating the shared parameters and treating them as if they are not common. We verified analytically that it does not matter which approach is taken since the hyperparameter $\alpha$ is entirely in one of the models at either $\alpha=0$ or $\alpha=1$.  In practice, when one choses to replicate the shared parameters so there are no overlapping parameters, then it turns out that the correct model still gets chosen but the posterior distributions of the parameters in the wrong model become very difficult to sample from and hence the autocorrelation time of the $\alpha$ chain is large, making it hard to accurately estimate the log-Bayes Factor. We therefore maintain the common parameters for both models which minimises the total number of parameters.

\subsection{Combined Likelihood Approach}
\label{sec:combined_likelihood_explanation}
The combined-likelihood method creates the supermodel by combining the two likelihoods via the hyperparameter $\alpha$. In this case, the two models are
completely distinctive, in the sense that the likelihood $\mathcal{L}_{1}$
and $\mathcal{L}_{2}$ only depend on the model parameters $\boldsymbol{\beta}$ and $\boldsymbol{\gamma}$ respectively. The
combined likelihood is then given by 

\begin{equation}
\mathcal{L}_{S}=f\left(\alpha\right)\mathcal{L}_{1}+\left(1-f\left(\alpha\right)\right)\mathcal{L}_{2}
\label{comblike}
\end{equation}

where $\mathcal{L}_{1}=\mathcal{P}\left(\mathcal{D}\left|\boldsymbol{\beta},\,\mathcal{M}_{1}\right.\right)$ and $\mathcal{L}_{2}=\mathcal{P}\left(\mathcal{D}\left|\boldsymbol{\gamma},\,\mathcal{M}_{2}\right.\right)$. If $f\left(\alpha\right)=\alpha$ the posterior probability distribution of $\alpha$ is obtained by marginalising over the parameters $\boldsymbol{\beta}$ and $\boldsymbol{\gamma}$ as follows,

\begin{equation}
\mathcal{P}\left(\alpha\left|\mathcal{D},\,\mathcal{M}_{1},\,\mathcal{M}_{2}\right.\right)\propto\int_{\boldsymbol{\gamma}}\int_{\boldsymbol{\beta}}\left[\alpha\mathcal{L}_{1}+\left(1-\alpha\right)\mathcal{L}_{2}\right]\mathcal{P}\left(\alpha,\,\boldsymbol{\beta},\,\boldsymbol{\gamma}\left|\mathcal{M}_{1},\,\mathcal{M}_{2}\right.\right)d\boldsymbol{\beta}d\boldsymbol{\gamma}
\end{equation}

The condition (\ref{eq:general_condition}) applies: setting $\alpha=1$ yields the Bayesian Evidence of model $\mathcal{M}_{1}$ while setting $\alpha=0$ gives
the Bayesian Evidence for model $\mathcal{M}_{2}$. If we assume the priors are separable, which is often the case, then we can write the above equation as 

\begin{equation}
\mathcal{P}\left(\alpha\left|\mathcal{D},\,\mathcal{M}_{1},\,\mathcal{M}_{2}\right.\right)\propto\mathcal{P}\left(\alpha\left|\mathcal{M}_{1},\,\mathcal{M}_{2}\right.\right)\int_{\boldsymbol{\gamma}}\int_{\boldsymbol{\beta}}\left[\alpha\left(\mathcal{L}_{1}-\mathcal{L}_{2}\right)+\mathcal{L}_{2}\right]\mathcal{P}\left(\boldsymbol{\beta},\,\boldsymbol{\gamma}\left|\mathcal{M}_{1},\,\mathcal{M}_{2}\right.\right)d\boldsymbol{\beta}d\boldsymbol{\gamma}
\label{linearpost}
\end{equation}

Since the above integration is independent of $\alpha$, the posterior will be of the form 
\[
\mathcal{P}\left(\alpha\left|\mathcal{D},\,\mathcal{M}_{1},\,\mathcal{M}_{2}\right.\right)\propto\mathcal{P}\left(\alpha\left|\mathcal{M}_{1},\,\mathcal{M}_{2}\right.\right)\left(m\alpha+c\right)
\]
If a flat or wide Gaussian distribution for the prior (centred
on $\alpha=0.5$ as we assume that $\mathcal{P}\left(\mathcal{M}_{i}\right)$
is equally likely) is imposed on $\alpha$ for $\alpha\in\left[0,1\right]$, 
then the posterior distribution of $\alpha$ is linear. Even in the case where the prior is not flat in $\alpha$ the key point is that it is analytically known. The
Bayes Factor, $B_{21}$ is then simply given by the ratio of the posterior
at the two endpoints. For a flat prior on $\alpha$ this gives: 

\begin{equation}
B_{21}=\dfrac{c}{m+c}
\end{equation}
where $m$ and $c$ are the constants derived from Eq. (\ref{linearpost}).  The posterior distribution of $\alpha$ needs to normalized, therefore we also have that $m = 2(1-c)$ 
and thus $B_{21} = \dfrac{c}{2-c}$. We now have a simple Bayesian parameter estimation problem, which is relatively straightforward to solve computationally using a Monte Carlo Markov Chain (MCMC) and a simple Metropolis-Hastings algorithm \citep{metropolis1953equation,hastings1970monte}. 

The fact that the posterior for $\alpha$ is simply a straight line greatly simplifies the determination of $B_{21}$ in practice when using a
sampling method. Since we know the functional form of $\PP(\alpha|\DD,\MM_1,\MM_2)$ we can use all MCMC samples to estimate  
the parameters $m$ and $c$, instead of only those where $\alpha\approx 0$
and $\alpha\approx1$. We can predict the accuracy with
which we can measure $B_{21}$ from $N$ MCMC samples that are distributed
with $\PP(\alpha|\DD,\MM_1,\MM_2)$ (the detailed calculation is shown in \ref{combined_likelihood_derivation}). We find that:

\begin{equation}
\sigma_{\textrm{log B}_{21}}\propto\dfrac{1}{\sqrt{N} c\left(2-c\right)}
\label{eq:precision}
\end{equation}

This shows that when $c$ or $(2-c)$ are small -- corresponding to very small or large Bayes factors -- accurate measurements require a large number of independent samples. Of course, one can argue that models that are disfavoured by a
larger Bayes factor do not require a very
accurate determination of $B_{21}$ to accurately perform model selection. Smaller Bayes factors can be determined more accurately with the same number of samples (refer to Figure (\ref{fig:mcmc_comp})).

\subsubsection{Sampling, Thinning and Convergence}
If we use a MCMC method to sample from the posterior, then, as we are dealing with either combined likelihoods or models, it is important to ensure the resulting MCMC samples are independent. Failure to ensure this leads to biases as the MCMC chain for $\alpha$ can be quite strongly autocorrelated. One way to reduce autocorrelation is to thin the chain, by recording only every $n_{\rm{Thin}}$ steps. We use the Python package \texttt{acor}\footnote{\texttt{\href{https://github.com/dfm/acor}{https://github.com/dfm/acor}}} to monitor the autocorrelation time of the chain. Smaller values of the autocorrelation length indicate less correlated samples, and hence that the chain has effectively more independent samples.An autocorrelation time of $\tau$ for $N$ samples effectively provides $\dfrac{N}{\tau}$ quasi-independent samples. In practice, we vary $n_{\rm{Thin}}$ until $\tau$ for the thinned chain is less than $10$. Empirically we find that this ensures unbiased parameter estimation, but can lead to $n_{\rm{Thin}}$ as large as $800$ in some of our runs.  In general ensure convergence of our MCMC chains by running all chains until they reach a Gelman-Rubin value of less than than $1.05$. 

\subsection{Combined Model Approach }
\label{sec:mixture_model_explanation}

We discussed above the implementation of the Supermodel idea through combining the models at the level of the likelihoods. Here we consider the alternative option: to combine them at the model level via a hyperparameter $\alpha$,
\begin{equation}
\MM_{S}=f\left(\alpha\right)\MM_{1}+\left(1-f\left(\alpha\right)\right)\MM_{2}
\label{combmod}
\end{equation}

In the case where $f\left(\alpha\right)=\alpha$, we will in this case usually assume a flat prior in the interval $[0,1]$ for $\alpha$, but other choices are possible (for example an enlarged interval which may make it easier to evaluate the posterior for $\alpha$ at $\alpha=0$ and $1$).
The posterior distribution of $\alpha$ is then given by 
\begin{equation}
\mathcal{P}\left(\alpha\left|\mathcal{D},\,\mathcal{M}_{1},\,\mathcal{M}_{2}\right.\right)=\int_{\boldsymbol{\gamma}}\int_{\boldsymbol{\beta}}\mathcal{P}\left(\mathcal{D}\left|\alpha,\,\boldsymbol{\beta},\,\boldsymbol{\gamma},\,\mathcal{M}_{1},\,\mathcal{M}_{2}\right.\right)\mathcal{P}\left(\alpha,\,\boldsymbol{\beta},\,\boldsymbol{\gamma}\left|\mathcal{M}_{1},\,\mathcal{M}_{2}\right.\right)d\boldsymbol{\beta}d\boldsymbol{\gamma}
\end{equation}

The objective is to find $\mathcal{P}\left(\alpha=0\left|\DD,\,\MM_{1},\,\MM_{2}\right.\right)$
and $\mathcal{P}\left(\alpha=1\left|\DD,\,\MM_{1},\,\MM_{2}\right.\right)$ because at these two endpoints, the posterior of $\alpha$ actually gives the Bayesian Evidence for each model. Hence, the Bayes Factor is given by 
\begin{equation}
B_{21}=\dfrac{\mathcal{P}\left(\alpha=0\left|\DD,\,\MM_{1},\,\MM_{2}\right.\right)}{\mathcal{P}\left(\alpha=1\left|\DD,\,\MM_{1},\,\MM_{2}\right.\right)}
\end{equation}

Although one can show analytically that this is correct, in practice, as we will explicitly show below, the marginal posterior of $\alpha$ can be  a complicated and unknown function of $\alpha$. We can obtain the Bayes factor only by considering samples with $\alpha\approx1$ and $\alpha\approx0$ which means we need to have a large number of samples in each limit. In contrast, the combined likelihood approach had the advantage that the posterior for $\alpha$ was simply a linear function, which makes it much easier and more accurate in practice to fit for the Bayes factor since all the samples can be used, as we will now demonstrate. 

\section{Application to Linear Model}
\label{sec:application_linear_model}
In this section, we apply the above supermodel methods to a simple
case study: the Gaussian linear model. This has the advantage that we can perform all calculations analytically, thus providing a benchmark for comparing our final results.

\subsection{Data}

\begin{figure}[t]
\noindent \begin{centering}
\includegraphics[width=12cm]{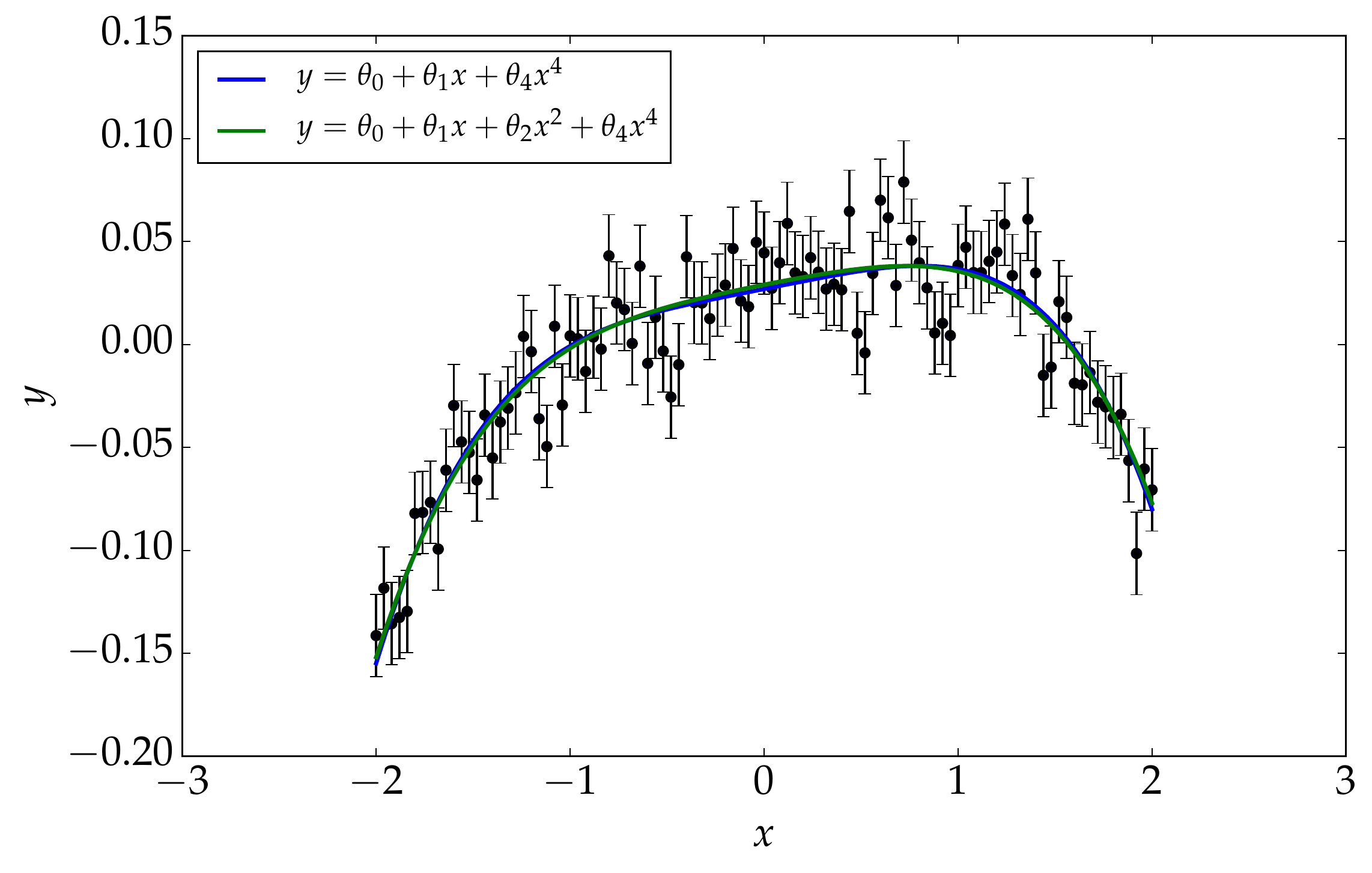}
\par\end{centering}
\caption{\textbf{Data generated from a fourth-order polynomial} -  The true model is the quartic without the quadratic term (thick blue line). The errors are normally distributed with $\sigma=0.02$. The Maximum a Posteriori (MAP) best-fits from the two models  $y=\theta_{0}+\theta_{1}x+\theta_{4}x^{4}$ and $y=\theta_{0}+\theta_{1}x+\theta_{2}x^{2}+\theta_{4}x^{4}$ are shown. Since the two fits are so similar it is not surprising that the simpler model has higher Bayesian evidence. }
\label{fig:lindat}
\end{figure}

In our toy model, the data (shown in Fig (\ref{fig:lindat})) has been generated from a fourth order
polynomial of the form $y=\theta_{0}+\theta_{1}x+\theta_{4}x^{4}$.
Since we know the correct model, it is easy to test if our methods
(as explained in the previous section) work successfully. 

Throughout this section, we wish to select between two models $\mathcal{M}_{1}$ and
$\mathcal{M}_{2}$ which are respectively given by 
\[
y=\theta_{0}+\theta_{1}x+\theta_{2}x^{2}+\theta_{4}x^{4}
\]
\[
y=\theta_{0}+\theta_{1}x+\theta_{4}x^{4}
\]
Since it is a linear model, we can calculate the Bayesian Evidence analytically (refer to \ref{BayesianEvidence}).
Due to the fact that $\mathcal{M}_{2}$ is nested in $\mathcal{M}_{1}$ at $\theta_{2}=0$, one can also compute the SDDR to verify the Bayes Factor calculated from the ratio of Bayesian Evidences (refer to \ref{BayesianEvidence}) is correct. 

The example data used here was actually generated from $\MM_2$. Since
$\MM_1$ is a model that contains $\MM_2$ (for $\theta_2=0$), it is not surprising
that the maximum log-likelihood of $\MM_2$ is higher than the one of $\MM_1$.
The model with the highest log-likelihood is the one with the most
freedom, $y=\theta_{0}+\theta_{1}x+\theta_{2}x^{2}+\theta_{3}x^{3}+\theta_{4}x^{4}$.
However, as expected, the model having the highest evidence is the model $\mathcal{M}_{2}$. This is an illustration of the Occam's Razor Effect, that
is, models with larger number of parameters will automatically be
penalised when calculating the Bayesian Evidence.

\subsection{Combined Likelihood}

Unlike the combined model method which has complexities (see Section \ref{sec:combined_model_linear}), the combined-likelihood method is relatively
simple. As explained in Section \ref{sec:combined_likelihood_explanation}, the posterior distribution of
the weight $\alpha$ is always linear: $\mathcal{P}\left(\alpha\left|\mathcal{D},\,\mathcal{M}_{1},\,\mathcal{M}_{2}\right.\right)=m\alpha+c$.
For a flat prior on the hyperparameter $\alpha$, one can then show that $m=2\left(1-c\right)$. 

\begin{figure}[htb]
\noindent \begin{centering}
\label{Figure 3}\includegraphics[width=12cm]{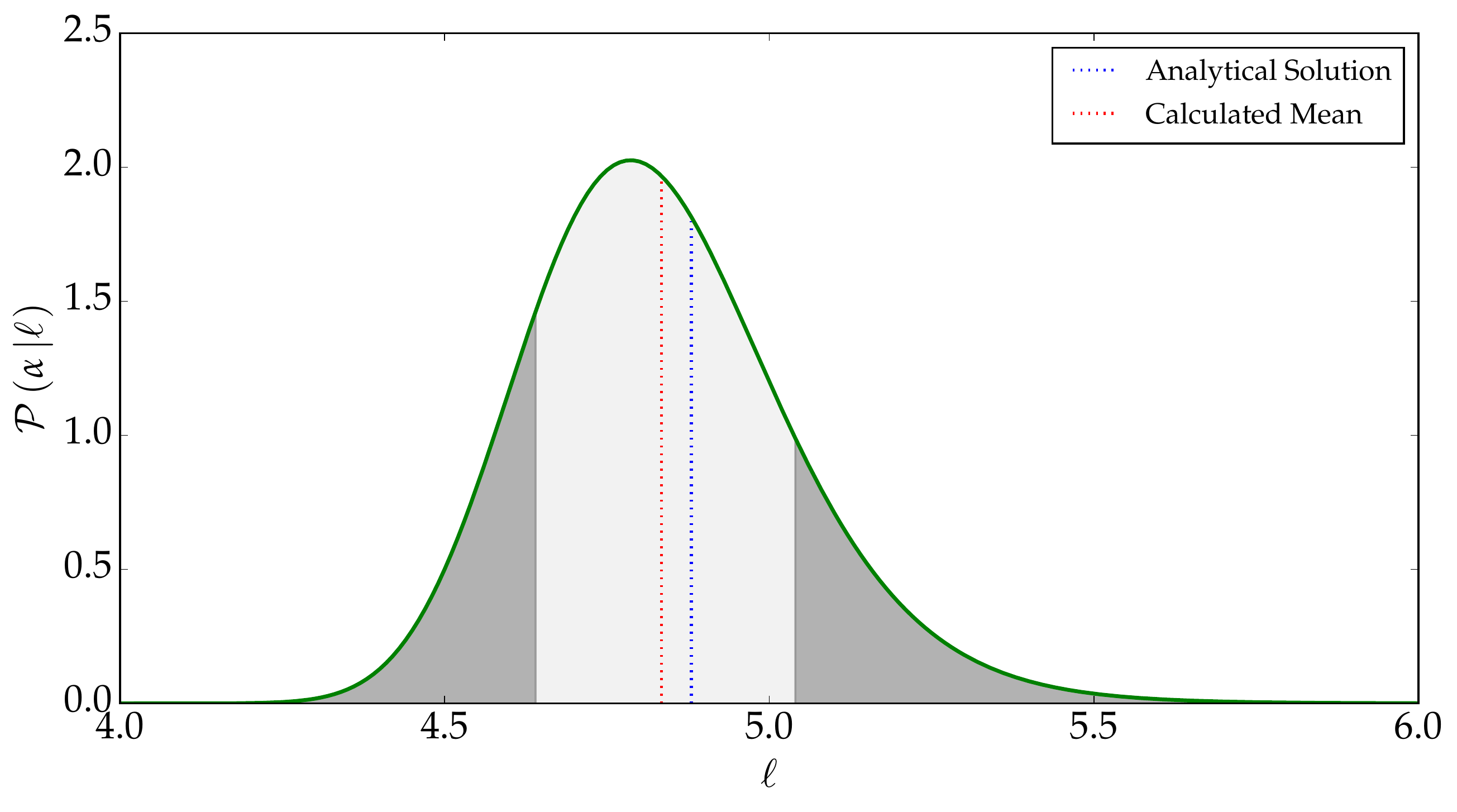}
\par\end{centering}
\caption{\textbf{Likelihood of the log-Bayes Factor} resulting from an MCMC with $2\times10^{7}$ steps and a thinning of 300, yielding around $66 000$ independent samples.The blue vertical dotted line
shows the analytical value while the red vertical dotted line shows
the value of $\textrm{log}\,B_{21}$ recovered from the MCMC samples.}
\label{fig:logbayes}
\end{figure}

Hence, the normalised posterior of $\alpha$ is given by 
\begin{equation}
\mathcal{P}\left(\alpha\left|\mathcal{D},\,\mathcal{M}_{1},\,\mathcal{M}_{2}\right.\right)=2\alpha\left(1-c\right)+c
\end{equation}
and
\[
\textrm{B}_{21}=\dfrac{\mathcal{P}\left(\alpha=0\left|\mathcal{D},\,\mathcal{M}_{1},\,\mathcal{M}_{2}\right.\right)}{\mathcal{P}\left(\alpha=1\left|\mathcal{D},\,\mathcal{M}_{1},\,\mathcal{M}_{2}\right.\right)}=\dfrac{c}{2-c}
\]
Setting $\ell=\textrm{log B}_{21}$ and under the assumption that the samples are uncorrelated, the likelihood of $\alpha$, 
$\mathcal{P}\left(\alpha\left|\ell\right.\right)$, is now given by: 
\begin{equation}
\mathcal{P}\left(\alpha\left|\ell\right.\right)=\prod_{i}\left[2\alpha_{i}+\dfrac{2\left(1-2\alpha_{i}\right)}{1+e^{-\ell}}\right]
\label{likefit}
\end{equation}

\begin{figure}[htb]
\noindent \begin{centering}
\includegraphics[width=12cm]{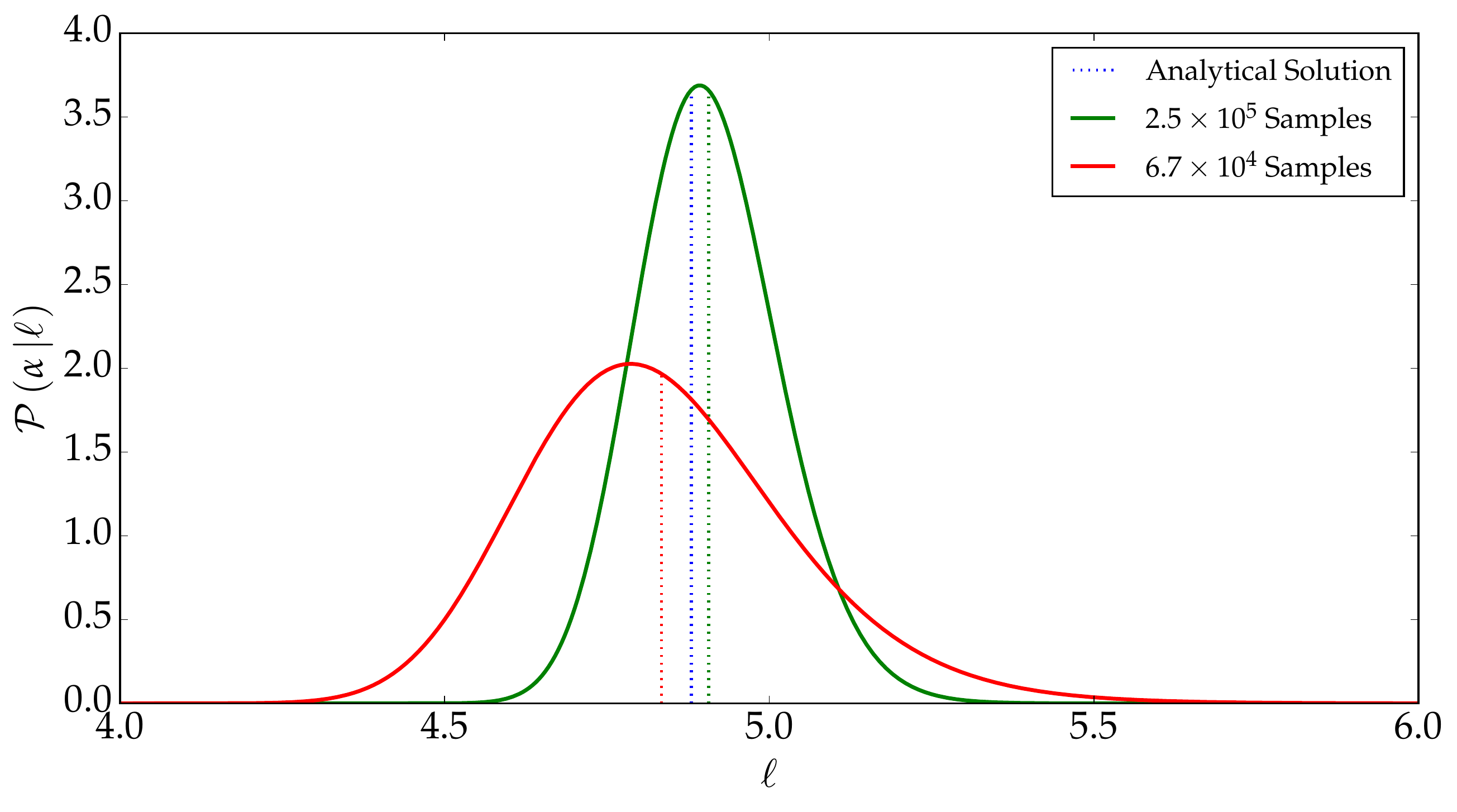}
\par\end{centering}
\caption{\textbf{Likelihood of the log-Bayes Factor for different number of samples} - As the number of samples increases, the precision with which the log-Bayes Factor is determined increases. In the above plot, $6.7\times10^{4}$ samples yields the a log-Bayes Factor estimate of $\textrm{log B}_{21}=4.83_{-0.19}^{+0.21}$ (red curve) while with roughly $2.5\times10^{5}$ samples, the log-Bayes Factor is estimated to be $\textrm{log B}_{21}=4.91\pm0.10$. In other words, in this case for the same log-Bayes Factor, while the number of samples has increased by roughly a factor of 4, the precision improves by a factor of 2, in agreement with Equation (\ref{eq:precision}).}
\label{fig:mcmc_comp}
\end{figure}

We tested that this approach works in practice by sampling
directly from the analytically-known distribution using Lahiri's method
(\citet{lahiri1951method} and \citet{book:408649}), and we found that
the result agreed with expectations. However, if the answer is not
already known then one can estimate it using MCMC. We use
the standard Metropolis-Hastings algorithm to obtain the $\alpha$
samples. Then, using Eq. (\ref{likefit}), we evaluate the likelihood $\mathcal{P}\left(\alpha\left|\ell\right.\right)$ on a grid of $\ell$ values. The result is shown in Fig (\ref{fig:logbayes}) with an estimated log-Bayes Factor of $\textrm{log B}_{21}=4.83_{-0.19}^{+0.21}$, which agrees well with the analytical result $\left(\textrm{log B}_{21}=4.88\right)$.

From the discussion in Section \ref{sec:combined_likelihood_explanation} and \ref{precision_bayes_factor} we know that we
need of order $10^5$ independent samples to determine the Bayes
factor sufficiently accurately which is comparable to the number of samples
needed in other methods like nested sampling \citep{skilling2006nested}.
Unfortunately the samples in a MCMC chain are correlated and we had to thin the chain
by a factor of 300 in order to obtain uncorrelated samples, implying that the method is significantly slower than nested sampling for this problem. This could be alleviated
by using other sampling methods to reduce the correlations,
for example Hamiltonian MC (HMC) \citep{neal2011mcmc}. The computational cost of
single HMC steps is however itself high unless one can compute
the gradients analytically. This situation changes for problems in very high dimensional
spaces relevant for many problems. Nested sampling scales exponentially with the number of model parameters while MCMC methods have a much better, polynomial, scaling. 

\subsection{Combined Model}
\label{sec:combined_model_linear}

In this section, we show that the combined model approach to supermodels also works, though we also consider its limitations. Refer to \ref{combined_model_derivation} for the analytical derivation of the posterior distribution of the hyperparameter $\alpha$ for the combined model. As explained in Section \ref{sec:mixture_model_explanation}, 
combined model is given by $\mathcal{M}_{3}=\alpha\mathcal{M}_{1}+\left(1-\alpha\right)\mathcal{M}_{2}$, and we need to estimate the posterior at the limits $\mathcal{P}\left(\alpha = 0\left|\mathcal{D},\,\mathcal{M}_{1},\,\mathcal{M}_{2}\right.\right)$ and $\mathcal{P}\left(\alpha=1\left|\mathcal{D},\,\mathcal{M}_{1},\,\mathcal{M}_{2}\right.\right)$ in order to calculate the Bayes Factor. This is tricky since it is difficult to get many samples near the boundaries $\alpha = 0,1$ and the resulting estimates are sensitive to binning artefacts and noise. 

\begin{figure}[H]
\noindent \begin{centering}
\includegraphics[width=11cm]{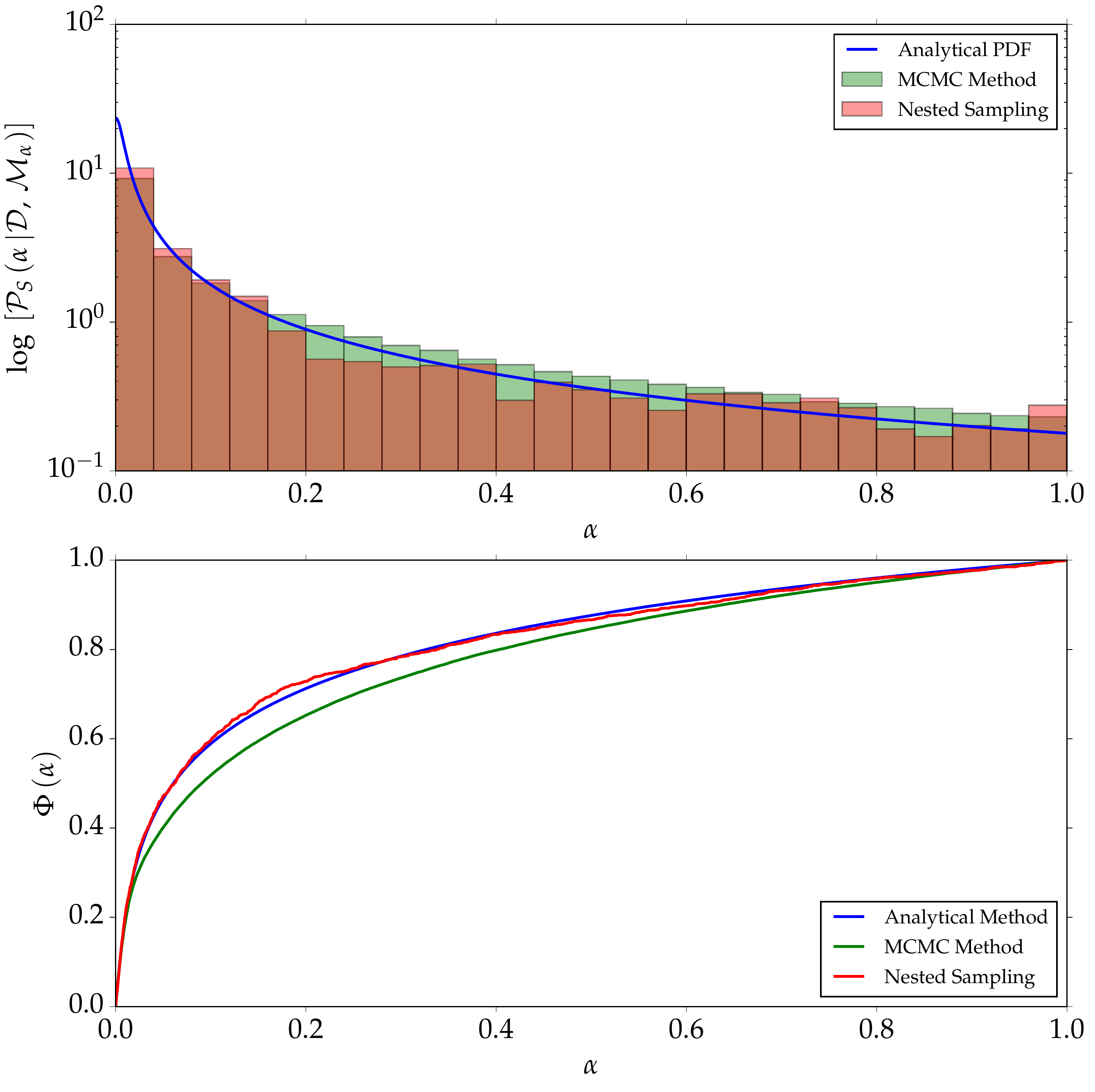}
\par\end{centering}
\caption{\textbf{Normalised Posterior and CDF of $\alpha$} - The top panel shows the normalised log-posterior distribution of $\alpha$ using three methods, analytical (shown in blue), MCMC (in green bins) and nested sampling (in red bins). Both nested sampling and MCMC perform badly at the boundaries (at $\alpha = 0$ and $\alpha=1$). Moreover, it is difficult to find a proper mathematical expression to fit for the posterior distribution. The bottom panel shows the cumulative distribution function (CDF), $\Phi\left(\alpha\right)$. Compared to MCMC, nested sampling performs better as it is well suited for dealing with multimodal distributions. However, we still have to deal with the issue of fitting the posterior.}
\label{fig:combined_model_alpha}
\end{figure}

While model selection in this case is highly accurate, the resulting estimates of the Bayes Factor are not highly accurate. The difficulty of sampling accurately in this model is shown in Fig (\ref{fig:combined_model_alpha}) where we show the normalised posterior and cumulative distribution functions for $\alpha$ estimated via MCMC and via nested sampling. The analytical results are also shown. The difficulty with this method is sampling efficiently at both boundaries. We consider an alternative, based on a reparameterization of $\alpha$, in Section \ref{sec:reparam_exploit}. Before that however, we consider application to a toy nonlinear problem. 

\section{Combined Likelihood Applied to Non-Linear Model}
\label{sec:application_non_linear}
We have demonstrated that the combined likelihood method works well for linear models. Here we explore its application to a toy nonlinear problem. 

\begin{figure}[!htb]
\noindent \begin{centering}
\includegraphics[width=12cm]{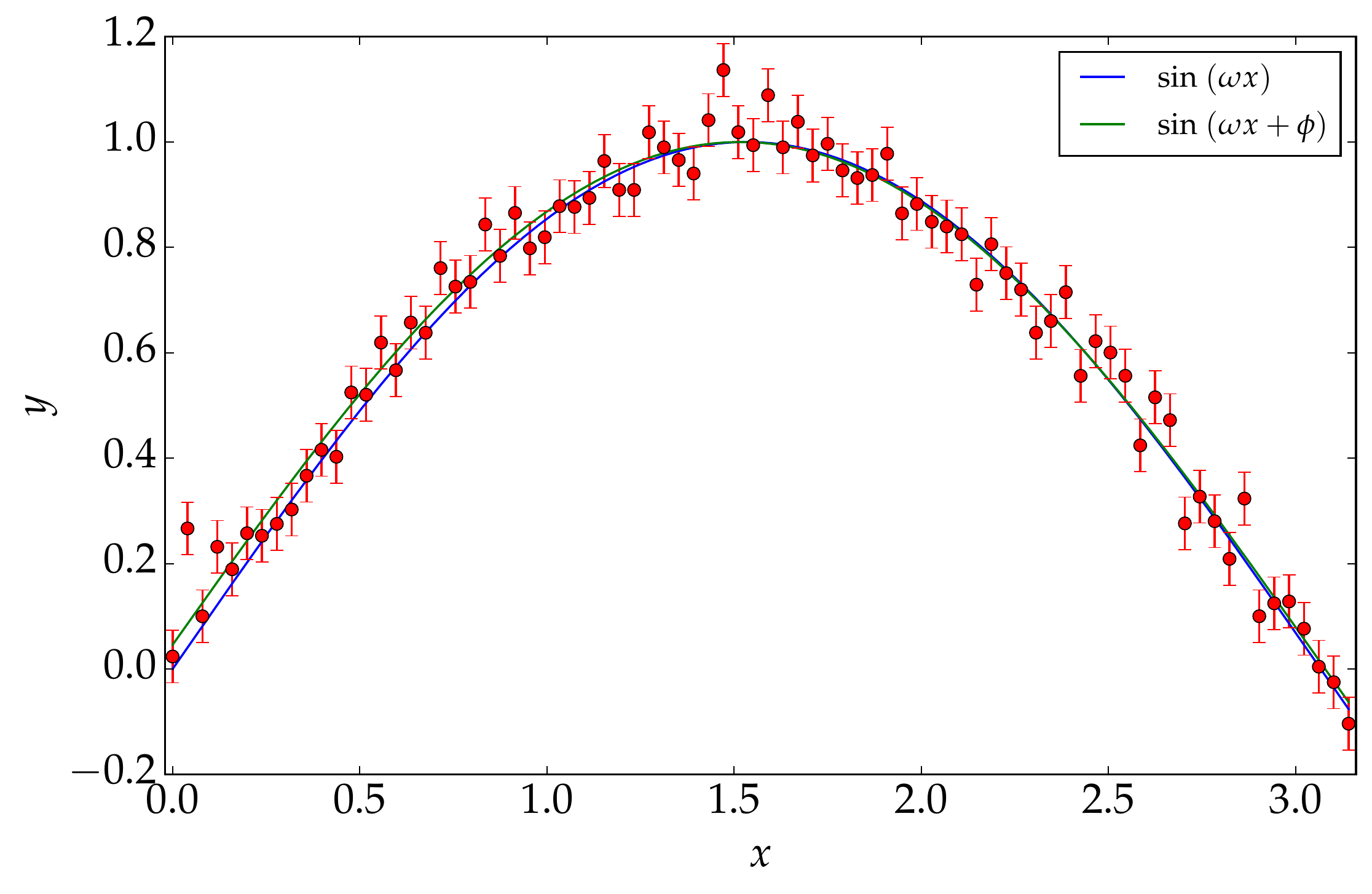}
\par\end{centering}
\caption{The toy nonlinear model we use to test the combined likelihood method. The two best individual model fits are shown. The data is generated from $\MM_2$ with $\omega = 1$ and $\phi = 0.06$.}
\label{fig:non_linear_data}
\end{figure}

We generate data from a sinusoidal function: 
\begin{equation}
y=\textrm{sin}\left(\omega x+\phi\right)
\end{equation}
$\MM_1$ contains just the parameter $\omega$ while $\MM_2$ contains both $\omega$ and the phase shift $\phi$. We add Gaussian noise with standard deviation $0.05$ and use fiducial values of $\omega=1.0$ and $\phi=0.06$ for $x\in\left[0,\,\pi\right]$ to generate the data from $\MM_2$ which is shown, along with the two best-fits from within $\MM_1$ and $\MM_2$ respectively, in Figure (\ref{fig:non_linear_data}). For the priors on $\omega$ and $\phi$ we choose independent Gaussians with $\mathcal{P}(\omega) \sim \mathcal{N}(1, 2^2)$ and $\mathcal{P}(\phi) \sim \mathcal{N}(0,0.05^2)$. 

Since the model is nonlinear we do not have an analytical solution for the Bayes Factor. Instead we use PyMultinest \citep{buchner2014x} to compute the Bayes Factor, finding $\textrm{log B}_{21}= 5.023\pm0.078$

\begin{figure}[!htb]
\noindent \begin{centering}
\includegraphics[width=12cm]{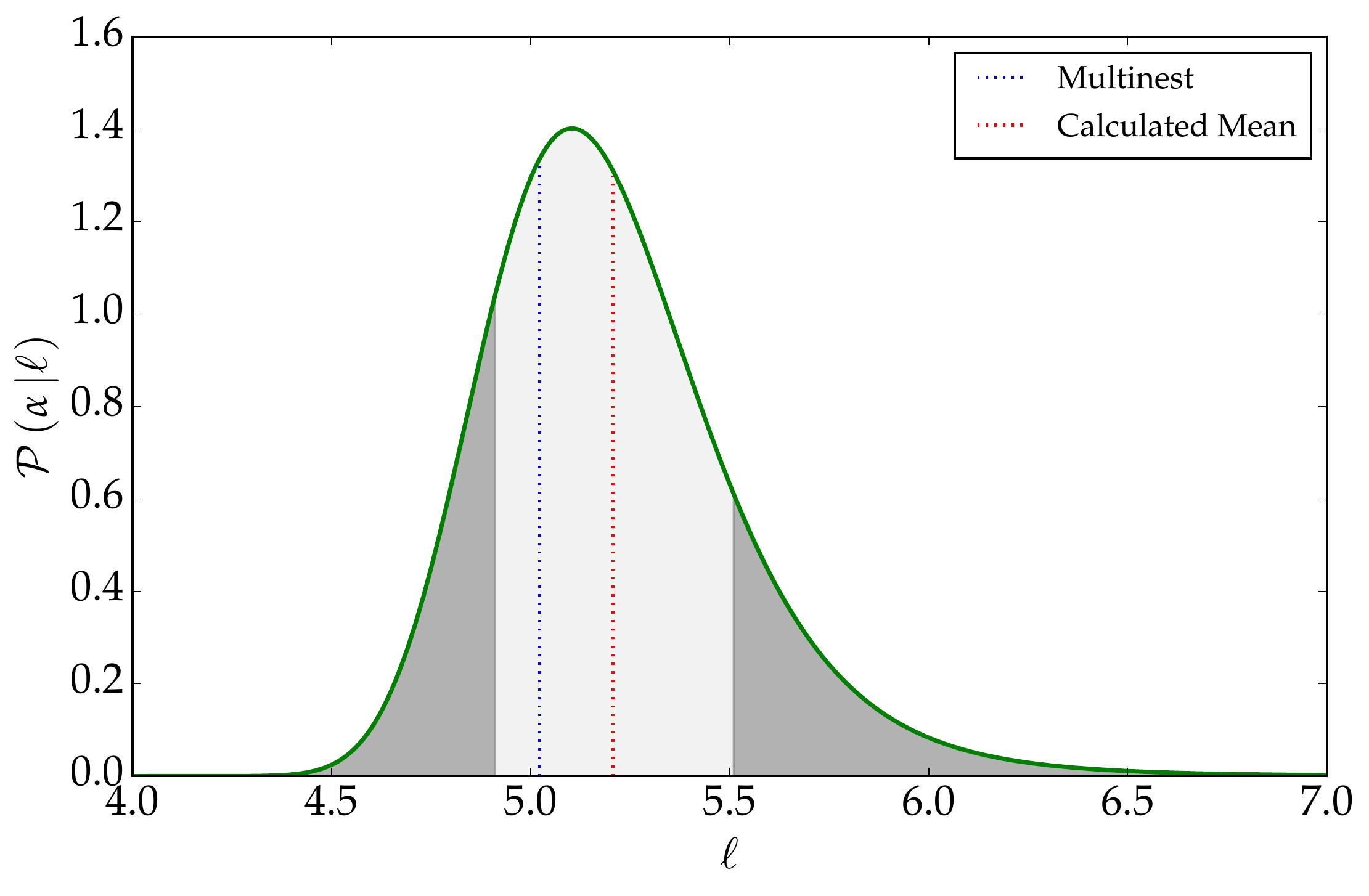}
\par\end{centering}
\caption{The inferred likelihood for $\alpha$ for the toy non-linear problem shown in Figure (\ref{fig:non_linear_data}). The resulting mean is fully consistent with the PyMultinest nested sampling result.}
\label{fig:likelihood_alpha_non_linear}
\end{figure}

Recall that in our method, the combined likelihood is 
\begin{equation}
\mathcal{L}=\alpha\mathcal{L}_{1}+\left(1-\alpha\right)\mathcal{L}_{2}
\end{equation}
We ran a chain of length $4 \times 10^7$ and use the appropriate thinning (in this case approximately 800) giving approximately $5 \times 10^4$ independent samples. The resulting distribution of the log-Bayes Factor is shown in Figure (\ref{fig:likelihood_alpha_non_linear}). The resulting mean of the log-Bayes Factor is given by $\textrm{log B}_{21}=5.21\pm0.30$, consistent with the PyMultiNest estimate of $\textrm{log B}_{21} = 5.023\pm0.078$. Of course, for such a small parameter space nested sampling is far superior in performance, however this gives evidence that the combined Likelihood method carries over successfully to nonlinear models.  

\section{Exploiting the Reparameterization of $\alpha$\label{sec:reparam_exploit}}

In this section, we test one possibility for the reparameterization freedom of $\alpha$ to deal with the challenges highlighted in Section \ref{sec:combined_model_linear}. In particular, we choose $\alpha\rightarrow e^{\alpha}$. We try both the combined model and the combined likelihood methods. 

\subsection{Combined Likelihood}

Assuming a flat prior for $\alpha$, its posterior distribution is now of the form 

\begin{equation}
\label{eq:repara_like}
\mathcal{P}\left(\alpha\left|\mathcal{D},\,\mathcal{M}_{1},\,\mathcal{M}_{2}\right.\right)=a\,e^{\alpha}+b
\end{equation}
where $\alpha \in (-\infty, 0]$.  Normalising this distribution requires a cutoff, $\Lambda$. We find that a cutoff of $\Lambda = -4$ enables a reliable estimate of the log-Bayes Factor. 

\begin{figure}[!htb]
\noindent \begin{centering}
\includegraphics[width=12cm]{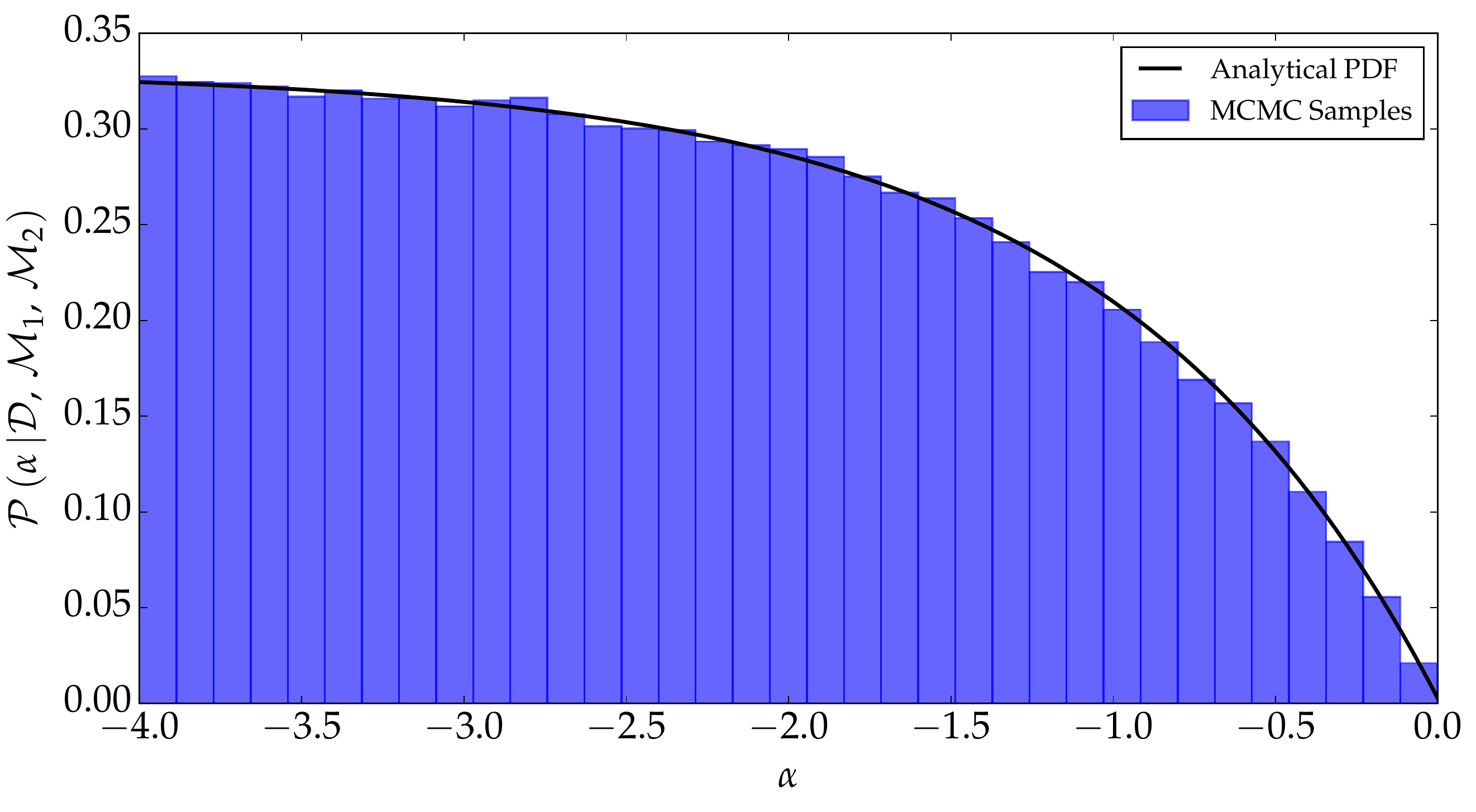}
\par\end{centering}
\caption{The analytical posterior distribution of $\alpha$ for the linear model is shown in black and the histogram corresponding to an MCMC run. The total number of steps in the MCMC is $5\times10^{6}$, the thinning factor was set to $15$ and eventually we have $\approx 330000$ recorded samples.}
\label{Figure 7}
\end{figure}

The ratio of the posterior of $\alpha$ at $\Lambda, 0$ estimates the Bayes Factor:

\[
\textrm{B}_{21}=\dfrac{\mathcal{P}\left(\alpha= \Lambda\left|\mathcal{D},\,\mathcal{M}_{1},\,\mathcal{M}_{2}\right.\right)}{\mathcal{P}\left(\alpha=0\left|\mathcal{D},\,\mathcal{M}_{1},\,\mathcal{M}_{2}\right.\right)} = \dfrac{ae^{\Lambda}+b}{a+b}
\]

\begin{figure}[!htb]
\noindent \begin{centering}
\includegraphics[width=12cm]{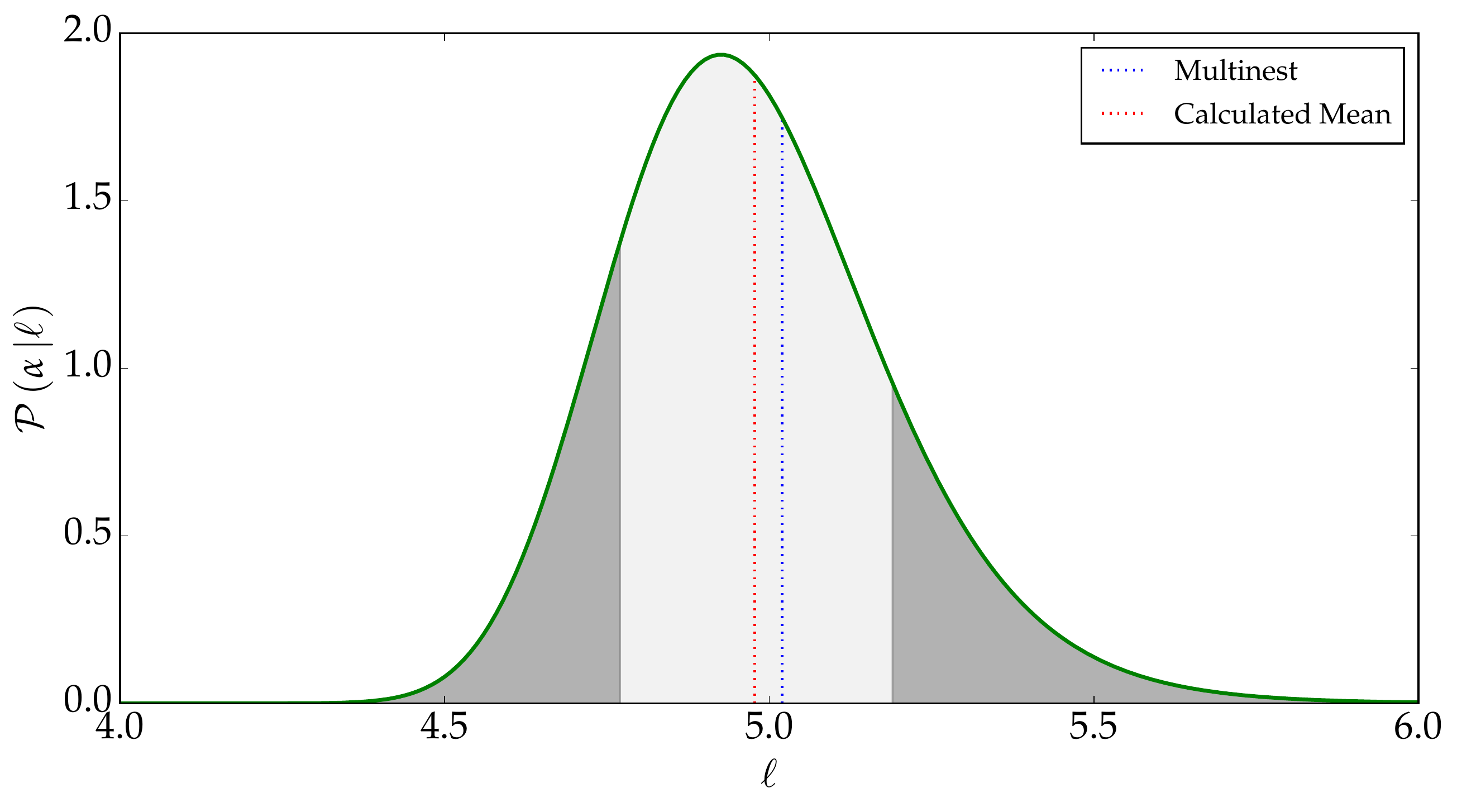}
\par\end{centering}

\caption{The likelihood of $\alpha$ given the log-Bayes Factor, $\ell$ for the non-linear models - Our result is consistent with the Multinest result at $1\sigma$ confidence interval. In the MCMC, the number of steps was fixed to $10^{7}$ with a thinning factor of 25, thus giving around $4\times10^{5}$ independent samples of $\alpha$.}
\end{figure}

Writing $\ell \equiv \textrm{log B}_{21}$ again we can now express $b$ in terms of $\ell$: 

\begin{equation}
b=\dfrac{1-e^{\Lambda-\ell}}{e^{-\ell}\left(\Lambda e^{\Lambda}+1-e^{\Lambda}\right)-\left(\Lambda+1-e^{\Lambda}\right)}
\end{equation}

The normalisation gives $a$ in terms of $b$ and $\Lambda$ and hence we can fit for the samples directly as we did earlier. We first try the method using the Linear model (which we used earlier) and since we can do everything analytically first, we can plot the posterior of $\alpha$; shown in Figure (\ref{Figure 7}). We fit for the $\alpha$ samples directly using Equation (\ref{eq:repara_like}) and our result is shown in the plot below.

\begin{figure}[!htb]
\noindent \begin{centering}
\includegraphics[width=12cm]{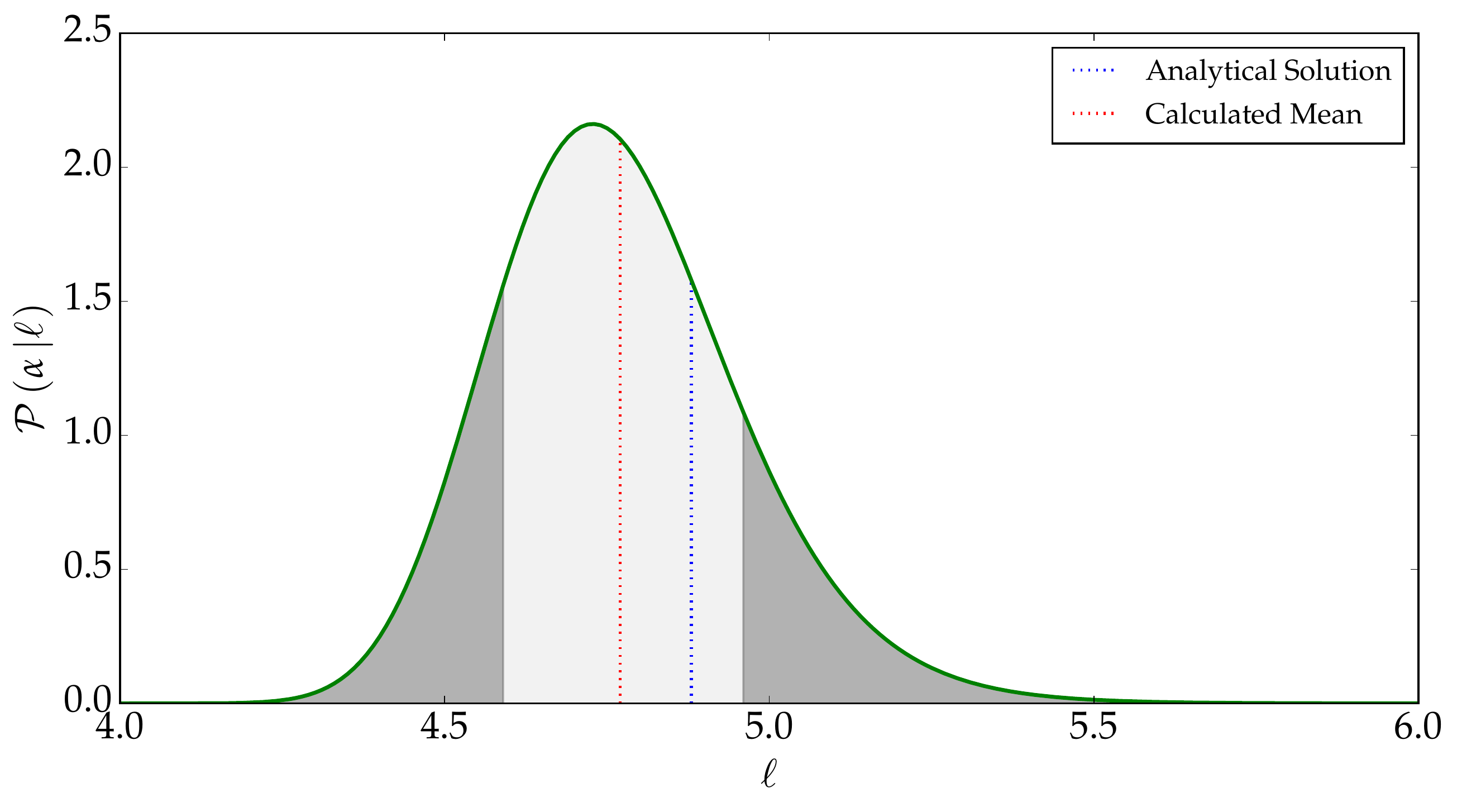}
\par\end{centering}
\caption{The likelihood of $\alpha$ as we vary $\ell \equiv \textrm{log B}_{21}$ in the linear models - The result is consistent with the log-Bayes factor which has been determined analytically.}
\end{figure}

The estimated log-Bayes Factor is given by: 

\begin{equation}
\textrm{log B}_{21}=4.77_{-0.18}^{+0.19}
\end{equation}
consistent with the analytically calculated log-Bayes Factor is $4.88$. 

\begin{figure}[htb]
\noindent \begin{centering}
\includegraphics[width=11cm]{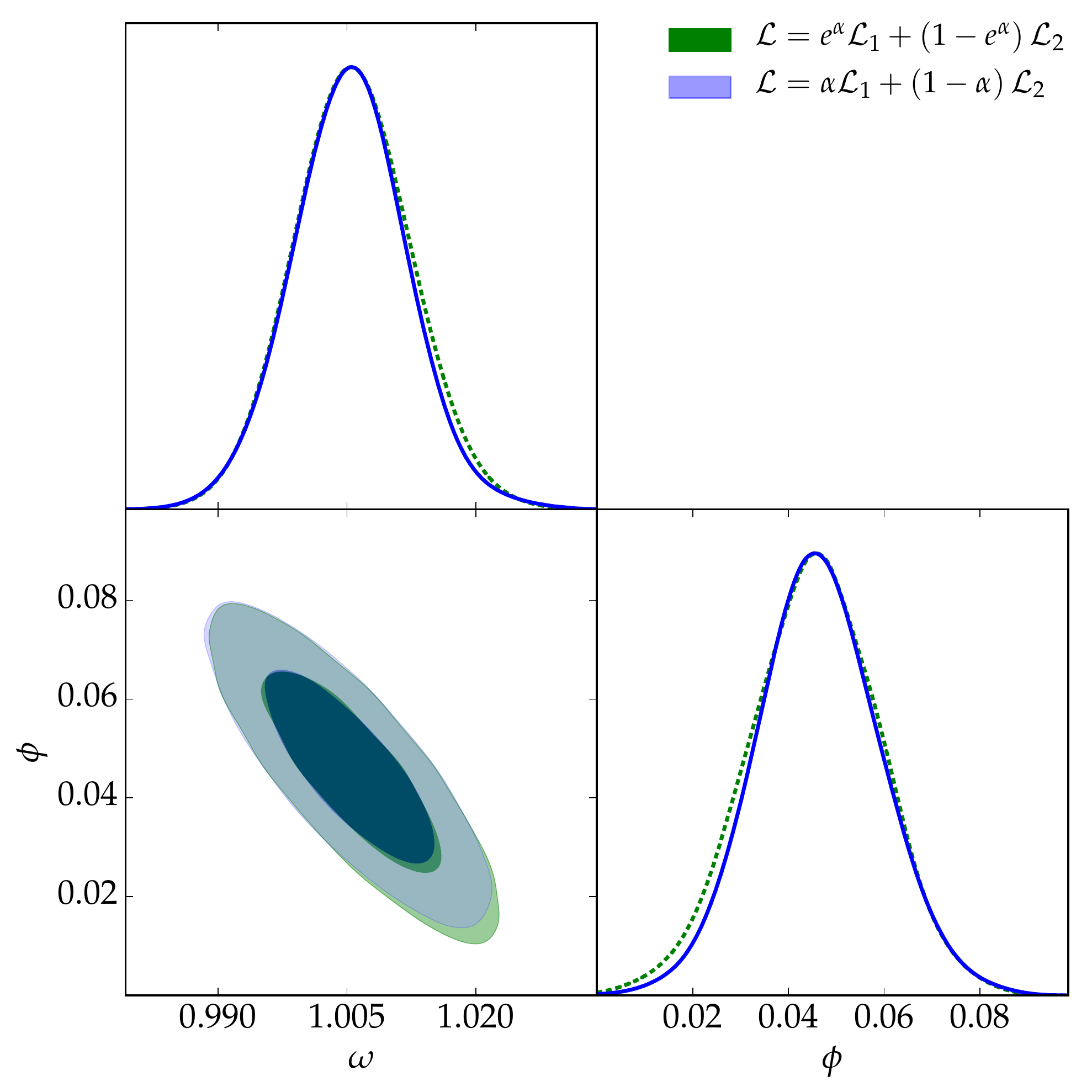}
\par\end{centering}
\caption{Comparing the posteriors of the nonlinear model parameters $\omega$ and $\phi$ for the combined likelihood method with and without reparametrisation of $\alpha$. 
The contours are essentially identical though because of the superior sampling properties of the $e^{\alpha}$ reparametrisation, the error on the log-Bayes Factor is reduced to $0.19$ compared to $0.3$.}
\end{figure}

Moreover, we can repeat the process with the non-linear model discussed earlier. The posterior distribution of the hyperparameter $\alpha$ still
follows Equation (\ref{eq:repara_like}) as we marginalise over
the models' parameters and not the hyperparameter $\alpha$.

In this case we find:
\begin{equation}
\textrm{log B}_{21}=4.98\pm0.19
\end{equation}

while  the log-Bayes Factor from Multinest is $5.023\pm0.078$. 

The advantage of using this transformation over the choice $f(\alpha) = \alpha$ is that  the sampling gets significantly better, requiring thinning factors of $\lesssim 50$. However, note that $\alpha\in\left[-4,\,0\right]$. One can decrease the lower limit of $\alpha$ further as the normalised posterior distribution of $\alpha$ follows the generic shape of the function $e^{\alpha}$, but would require many more samples. Therefore, in short, this method is significantly less computationally expensive but we still have to find the trade-off between the number of samples and the thinning factor.  The net result is that the error on $\textrm{log B}_{21}$ is reduced from $0.3$ to $0.19$ with no change in model parameter posteriors as shown in the above figure. 

\subsection{Combined Model}

Now let us study the performance of the combined model, given by 

\begin{equation}
\mathcal{M}_{3}=e^{\alpha}\mathcal{M}_{1}+\left(1-e^{\alpha}\right)\mathcal{M}_{2}
\end{equation}

\begin{figure}[H]
\noindent \begin{centering}
\includegraphics[width=12cm]{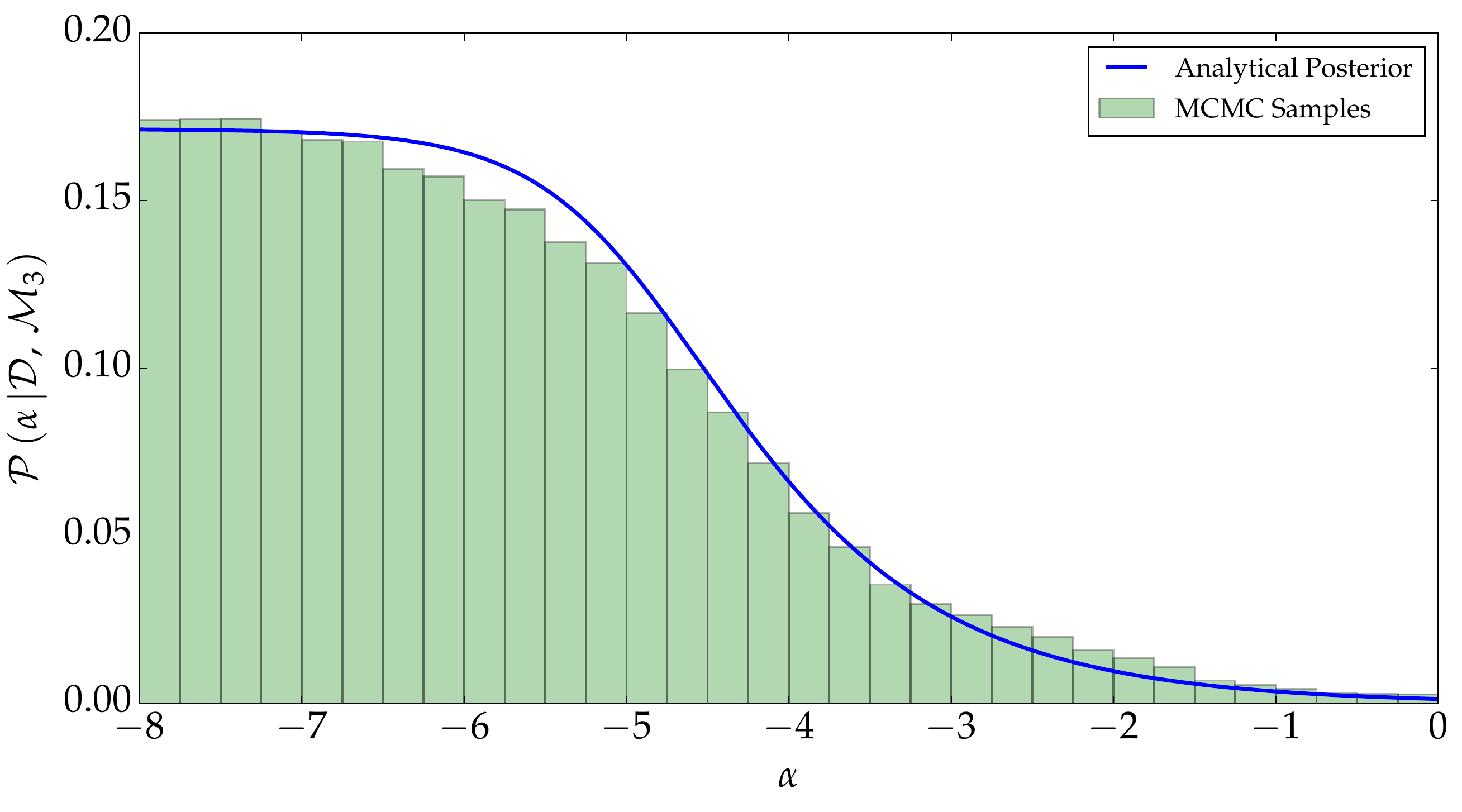}
\par\end{centering}
\caption{The normalised posterior distribution of $\alpha$ in the combined model for the linear model data - The number of samples is $\sim 2 \times 10^{5}$, followed by fixing the number of iterations in the MCMC to $2\times10^{6}$, with a thinning factor of 10. The curve in blue shows the analytical posterior distribution. Fitting the functional form $a\,\textrm{tanh}\left(b\alpha+c\right)+d$ to the histogram, where the parameters $a,\,b,\,c\textrm{ and }d$ are determined via optimisation, leads to a log-Bayes Factor of 3.59 instead of the true value of 4.88, primarily due to sampling issues.}
\label{combined_model_exp}
\end{figure}

The posterior as shown in Figure (\ref{combined_model_exp}) is now better behaved compared the previous combined model with $f(\alpha) = \alpha$ (refer to Figure (\ref{fig:combined_model_alpha})) although it is still not perfect, as evident from the differences between the analytical fit and the MCMC histogram. It is also easier to sample from the posterior distribution with $f(\alpha) = e^\alpha$. One can try to fit the resulting normalised histogram generated from the MCMC to a guessed functional form such as $a\,\textrm{tanh}\left(b\alpha+c\right)+d$ where $a,\,b,\,c,\textrm{and }d$ are the new parameters to be determined. Unfortunately we have no theoretical guidance as to the true analytical function to use and hence the recovered Bayes Factor is susceptible to systematic errors due to incorrect choice of function to be fitted. As a result, the combined likelihood approach appears superior.

\section{Summary and Conclusion}
\label{sec:conclusion}

In this work we have used the Savage-Dickey Density Ratio (SDDR) to show
that we can calculate the Bayes Factor of two non-nested models by introducing
a new hyperparameter that combines the models into a single supermodel. This Savage-Dickey Supermodel (SDSM) method does not need the Bayesian evidence (Marginal Likelihood) to be computed. The core supermodel embedding can be done either at the level of the model (eq. (\ref{combmod})) or at the level of the likelihood (eq.  (\ref{comblike})) and effectively makes the the models nested and hence amenable to the SDDR approach to computing the Bayes Factors. In the context of Gaussian linear models we show that the SDDR both analytically and numerically reproduces the Bayes Factors computed analytically. We then consider a nonlinear example and show that our supermodel approach agrees well with that from nested sampling. 

Though we have a clever way of avoiding multidimensional integrals
to calculate the Bayesian Evidence, this new method requires very
efficient sampling and for a small number of dimensions is not faster than individual
nested sampling runs. The major reason for this is that we require independent samples
for $\alpha$ and one way to ensure we are doing so is to have a short
autocorrelation length. Hence the thinning factor for the MCMC chain
needs to be adjusted as well as the number of the steps, especially
for large log-Bayes Factor. However, generically the scaling
of MCMC methods with the number of dimensions is much more benign
than the scaling of nested sampling methods. The approach presented
here is thus expected to work also for very high numbers of dimensions
where nested sampling fails. Additionally, if we only keep in a MCMC
chain the elements for which $\alpha=1$ or $\alpha=0$ then we obtain
 a model-averaged posterior. For this application we do not need
a very high number of samples, so that the method is competitive
with nested sampling for model averaged posteriors also at a smaller number of dimensions.

For future work we note that other, nonlinear, combinations of models/likelihoods are also possible. For example, consider product combined model and likelihood
$\mathcal{M}_{3}=\mathcal{M}_{1}^{\alpha}\mathcal{M}_{2}^{\left(1-\alpha\right)}$
and $\mathcal{L}_{3}=\mathcal{L}_{1}^{\alpha}\mathcal{L}_{2}^{\left(1-\alpha\right)}$
in which case, the general condition (\ref{eq:general_condition})
still holds for $\alpha\in\left[0,\,1\right]$. 

Such nonlinear supermodels, choices of reparametrisation function $f\left(\alpha\right)$ or other innovations (such as using simulated annealing) may greatly simplify some aspects of the sampling and provide a clever way of not only obtaining the log-Bayes Factor, which helps us to understand the relative strength of the models but also to have model averaged posteriors of all the parameters in both models.
Study of these generalisations is left to future work.

%% Numbered

%% Numbered without titles
%\bibliographystyle{model1a-num-names}

%% Harvard
% \bibliographystyle{model2-names.bst}\biboptions{authoryear}

%% Vancouver numbered
%\usepackage{numcompress}\bibliographystyle{model3-num-names}

%% Vancouver name/year
%\usepackage{numcompress}\bibliographystyle{model4-names}\biboptions{authoryear}

%% APA style
%\bibliographystyle{model5-names}\biboptions{authoryear}

%% AMA style
% \usepackage{numcompress}\bibliographystyle{model6-num-names}

\bibliographystyle{model1a-num-names}\biboptions{authoryear}
%\section*{\refname}
\bibliography{ref-1}

\newpage
\appendix
%\section*{Appendix A - The Gaussian Linear Model}
\section{Bayesian Evidence and SDDR for Gaussian Linear Models}
\label{BayesianEvidence}

Consider a polynomial of order $n-1$, that is, 

\[
y=\theta_{0}+\theta_{1}x+\theta_{2}x^{2}+\ldots+\theta_{n-1}x^{n-1}
\]
This model can be written in a general
form as 

\[
y=\sum_{k=0}^{n-1}\theta_{k}X_{k}
\]
or equivalently in matrix format as 

\[
\mathbf{y}=\mathbf{X}\boldsymbol{\theta}
\]
where $X_{0},\,X_{1},\ldots X_{n-1}$ are known as the basis functions.
If the measurement error $\sigma_{i}$ is known for each data point,
then we can define the design matrix as 

\[
D_{ij}=\dfrac{X_{j}\left(x_{i}\right)}{\sigma_{i}}
\]

Let us first derive the Bayesian Evidence, $\mathbb{Z}$, for such models. In matrix
format, we can write the prior as 

\[
\mathcal{P}\left(\boldsymbol{\theta}\left|\mathcal{M}\right.\right)=\dfrac{1}{\sqrt{\left|2\pi\mathbf{\mathbf{P}^{-1}}\right|}}\,\textrm{exp}\left(-\dfrac{1}{2}\boldsymbol{\theta}^{\textrm{T}}\mathbf{P}^{-1}\boldsymbol{\theta}\right)
\]

where $\mathbf{\mathbf{P}^{-1}}$ is the inverse of the covariance
matrix for the priors. The likelihood is given by 

\[
\mathcal{P}\left(\mathcal{D}\left|\boldsymbol{\theta},\mathcal{M}\right.\right)=\dfrac{1}{{\displaystyle \prod_{i}}\sqrt{\left(2\pi\sigma_{i}\right)}}\,\textrm{exp}\left[-\dfrac{1}{2}\left(\mathbf{b}-\mathbf{D}\boldsymbol{\theta}\right)^{\textrm{T}}\left(\mathbf{b}-\mathbf{D}\boldsymbol{\theta}\right)\right]
\]

where $\mathbf{b}$ is the vector $\left(\dfrac{y_{0}}{\sigma_{0}},\,\dfrac{y_{1}}{\sigma_{1}},\ldots\dfrac{y_{N-1}}{\sigma_{N-1}}\right)$
and $\mathbf{D}$ is the design matrix. The Bayesian Evidence, $\mathbb{Z}$
is then given by 

\[
\mathbb{Z}=\int\mathcal{P}\left(\mathcal{D}\left|\boldsymbol{\theta},\mathcal{M}\right.\right)\,\mathcal{P}\left(\boldsymbol{\theta}\left|\mathcal{M}\right.\right)\,d\boldsymbol{\theta}
\]

\[
\mathbb{Z}=\dfrac{1}{{\displaystyle \prod_{i}}\sqrt{\left(2\pi\sigma_{i}\right)}}\,\dfrac{\mathbf{b}^{\textrm{T}}\mathbf{b}}{\left|2\pi\mathbf{P}^{-1}\right|}\int\textrm{exp}\left[-\dfrac{1}{2}\left\{ \boldsymbol{\theta}^{\textrm{T}}\left(\mathbf{D}^{\textrm{T}}\mathbf{D}+\mathbf{\mathbf{P}^{-1}}\right)\boldsymbol{\theta}-2\boldsymbol{\theta}^{\textrm{T}}\mathbf{D}^{\textrm{T}}\mathbf{b}\right\} \right]d\boldsymbol{\theta}
\]

If we have a quadratic expression such as $\mathbf{x}^{\textrm{T}}\mathbf{A}\mathbf{x}+\mathbf{x}^{\textrm{T}}\mathbf{b}+\mathbf{c}$,
then this can be expressed as 

\[
\left(\mathbf{x}-\mathbf{h}\right)^{\textrm{T}}\mathbf{A}\left(\mathbf{x}-\mathbf{h}\right)+\mathbf{k}
\]

where 

\[
\mathbf{h}=-\dfrac{1}{2}\mathbf{A}^{-1}\mathbf{h}
\]

\[
\mathbf{k}=\mathbf{c}-\dfrac{1}{4}\mathbf{b}^{\textrm{T}}\mathbf{A}^{-1}\mathbf{b}
\]

Therefore, 

\[
\mathbb{Z}=\dfrac{1}{{\displaystyle \prod_{i}}\sqrt{\left(2\pi\sigma_{i}\right)}}\,\dfrac{\mathbf{b}^{\textrm{T}}\mathbf{b}\,\textrm{exp}\left(-\dfrac{1}{2}\mathbf{k}\right)}{\left|2\pi\mathbf{\mathbf{P}^{-1}}\right|}\int\textrm{exp}\left[-\dfrac{1}{2}\left(\boldsymbol{\theta-\mathbf{h}}\right)^{\textrm{T}}\left(\mathbf{D}^{\textrm{T}}\mathbf{D}+\mathbf{\mathbf{P}^{-1}}\right)\left(\boldsymbol{\theta-\mathbf{h}}\right)\right]d\boldsymbol{\theta}
\]

where 

\[
\mathbf{k}=-\left(\mathbf{D}^{\textrm{T}}\mathbf{b}\right)^{\textrm{T}}\left(\mathbf{D}^{\textrm{T}}\mathbf{D}+\mathbf{\mathbf{P}^{-1}}\right)^{-1}\left(\mathbf{D}^{\textrm{T}}\mathbf{b}\right)\hspace{1.5cm}\mathbf{h}=\left(\mathbf{D}^{\textrm{T}}\mathbf{D}+\mathbf{\mathbf{P}^{-1}}\right)^{-1}\left(\mathbf{D}^{\textrm{T}}\mathbf{b}\right)
\]

\[
\mathbb{Z}=\dfrac{\mathbf{b}^{\textrm{T}}\mathbf{b}\,\textrm{exp}\left(-\dfrac{1}{2}\mathbf{k}\right)}{{\displaystyle \prod_{i}}\sqrt{\left(2\pi\sigma_{i}\right)}}\,\sqrt{\dfrac{\left|2\pi\left(\mathbf{D}^{\textrm{T}}\mathbf{D}+\mathbf{\mathbf{\mathbf{P}^{-1}}}\right)^{-1}\right|}{\left|2\pi\mathbf{\mathbf{\mathbf{P}^{-1}}}\right|}}
\]

In particular, in this paper we will assume the prior on the each parameter is an independent Gaussian
centred on 0 with standard deviation equal to 1, and hence $\mathbf{\mathbf{\mathbf{P}^{-1}}}=\mathbf{I}$. We now derive the SDDR in the case when one model is nested in another.
Consider the two models $\mathcal{M}_{1}$ and $\mathcal{M}_{2}$
which are given by $y=\theta_{0}+\theta_{1}x+\theta_{2}x^{2}+\theta_{4}x^{4}$
and $y=\theta_{0}+\theta_{1}x+\theta_{4}x^{4}$ respectively. If we
define $\boldsymbol{\phi}_{1}=\left(\theta_{0},\,\theta_{1},\,\theta_{4}\right)$
and $\boldsymbol{\phi}_{2}=\left(\theta_{2}\right)$, we can then
write the likelihood and the priors as 

\[
\mathcal{P}\left(\mathcal{D}\left|\mathcal{M}_{1},\,\boldsymbol{\phi}_{1},\,\boldsymbol{\phi_{2}}\right.\right)=\dfrac{1}{{\displaystyle \prod_{i}}\sqrt{\left(2\pi\sigma_{i}\right)}}\,\textrm{exp}\left[-\dfrac{1}{2}\left(\mathbf{b}-\mathbf{D}_{1}\boldsymbol{\phi_{1}-\mathbf{D}_{2}\boldsymbol{\phi}_{2}}\right)^{\textrm{T}}\left(\mathbf{b}-\mathbf{D}_{1}\boldsymbol{\phi_{1}-\mathbf{D}_{2}\boldsymbol{\phi}_{2}}\right)\right]
\]

\[
\mathcal{P}\left(\boldsymbol{\phi}_{1}\left|\mathcal{M}_{1}\right.\right)=\dfrac{1}{\sqrt{\left|2\pi\mathbf{C}_{1}^{-1}\right|}}\,\textrm{exp}\left(-\dfrac{1}{2}\boldsymbol{\phi}_{1}^{\textrm{T}}\mathbf{C}_{1}^{-1}\boldsymbol{\phi}_{1}\right)\hspace{1.5cm}\mathcal{P}\left(\boldsymbol{\phi}_{2}\left|\mathcal{M}_{1}\right.\right)=\dfrac{1}{\sqrt{\left|2\pi\mathbf{C}_{2}^{-1}\right|}}\,\textrm{exp}\left(-\dfrac{1}{2}\boldsymbol{\phi}_{2}^{\textrm{T}}\mathbf{C}_{2}^{-1}\boldsymbol{\phi}_{2}\right)
\]

where $\mathbf{C}_{1}$ and $\mathbf{C}_{2}$ are covariance matrices
of size 3 and 1 respectively and $\mathbf{D}_{1}$ and $\mathbf{D}_{2}$
are the appropriate design matrices. In this case, $\mathbf{C}_{1}^{-1}=\mathbf{I}_{1}$
and $\mathbf{C}_{2}^{-1}=\mathbf{I}_{2}$ are in fact the Fisher Information
matrix of $\boldsymbol{\phi}_{1}$ and $\boldsymbol{\phi}_{2}$ respectively.
The SDDR is given by 

\[
\textrm{SDDR}=\left.\dfrac{\mathcal{P}\left(\boldsymbol{\phi}_{2}\left|\mathcal{D},\,\mathcal{M}_{1}\right.\right)}{\mathcal{P}\left(\boldsymbol{\phi}_{2}\left|\mathcal{M}_{1}\right.\right)}\right|_{\boldsymbol{\phi}_{2}=0}
\]

Therefore,

\[
\mathcal{P}\left(\boldsymbol{\phi}_{2}\left|\mathcal{D},\,\mathcal{M}_{1}\right.\right)\propto\mathcal{P}\left(\boldsymbol{\phi}_{2}\left|\mathcal{M}_{1}\right.\right)\int\mathcal{P}\left(\mathcal{D}\left|\mathcal{M}_{1},\,\boldsymbol{\phi}_{1},\,\boldsymbol{\phi_{2}}\right.\right)\mathcal{P}\left(\boldsymbol{\phi}_{1}\left|\mathcal{M}_{1}\right.\right)\,d\boldsymbol{\phi}_{1}
\]

The normalised posterior distribution of $\boldsymbol{\phi}_{2}$
is given by 

\[
\mathcal{P}\left(\boldsymbol{\phi}_{2}\left|\mathcal{D},\,\mathcal{M}_{1}\right.\right)=\dfrac{1}{\sqrt{\left|2\pi\mathbf{B}^{-1}\right|}}\,\textrm{exp}\left[-\dfrac{1}{2}\left(\boldsymbol{\phi}_{2}^{\textrm{T}}\mathbf{B}\boldsymbol{\phi}_{2}+2\boldsymbol{\phi}_{2}^{\textrm{T}}\mathbf{E}-\mathbf{E}^{\textrm{T}}\mathbf{B}^{-1}\mathbf{E}\right)\right]
\]

where 

\[
\mathbf{A}=\left(\mathbf{D}_{1}^{\textrm{T}}\mathbf{D}_{1}+\mathbf{C}_{1}^{-1}\right)^{-1}
\]

\[
\mathbf{B}=\mathbf{C}_{2}^{-1}+\mathbf{D}_{2}^{\textrm{T}}\mathbf{D}_{2}-\mathbf{D}_{2}^{\textrm{T}}\mathbf{D}_{1}\mathbf{A}\mathbf{D}_{1}^{\textrm{T}}\mathbf{D}_{2}
\]

\[
\mathbf{E}=\mathbf{D}_{2}^{\textrm{T}}\mathbf{D}_{1}\mathbf{A}\mathbf{D}_{1}^{\textrm{T}}\mathbf{b}-\mathbf{D}_{2}\mathbf{b}
\]

Then,

\[
\textrm{SDDR}=\sqrt{\dfrac{\left|2\pi\mathbf{C}_{2}^{-1}\right|}{\left|2\pi\mathbf{B}^{-1}\right|}}\,\textrm{exp}\left(-\dfrac{1}{2}\mathbf{E}^{\textrm{T}}\mathbf{B}^{-1}\mathbf{E}\right)
\]

\section{Combined Model - linear model}
\label{combined_model_derivation}

In this case, the two models are nested as 

\[
\mathcal{M}_{3}=\alpha\mathcal{M}_{1}+\left(1-\alpha\right)\mathcal{M}_{2}
\]

With the two models used in the text, the mixture model is 
written as 

\[
\mathcal{M}_{3}=\alpha\theta_{2}x^{2}+\theta_{0}+\theta_{1}x+\theta_{4}x^{4}
\]

Hence, the likelihood of the mixture model can be written as 

\[
\mathcal{P}\left(\mathcal{D}\left|\mathcal{M}_{3},\,\boldsymbol{\phi}_{1},\,\boldsymbol{\phi_{2}},\,\alpha\right.\right)=\dfrac{1}{{\displaystyle \prod_{i}}\sqrt{\left(2\pi\sigma_{i}\right)}}\,\textrm{exp}\left[-\dfrac{1}{2}\left(\mathbf{b}-\mathbf{D}_{1}\boldsymbol{\phi_{1}-\alpha\mathbf{D}_{2}\boldsymbol{\phi}_{2}}\right)^{\textrm{T}}\left(\mathbf{b}-\mathbf{D}_{1}\boldsymbol{\phi_{1}-\alpha\mathbf{D}_{2}\boldsymbol{\phi}_{2}}\right)\right]
\]

where $\mathbf{D}_{1}$ and $\mathbf{D}_{2}$
are the appropriate design matrices, as before.

The posterior distribution of $\alpha$ is then given by 

\[
\mathcal{P}\left(\alpha\left|\mathcal{D},\,\mathcal{M}_{3}\right.\right)\propto\int_{\boldsymbol{\phi}_{2}}\int_{\boldsymbol{\phi}_{1}}\mathcal{P}\left(\mathcal{D}\left|\mathcal{M}_{3},\,\boldsymbol{\phi}_{1},\,\boldsymbol{\phi_{2}},\,\alpha\right.\right)\mathcal{P}\left(\boldsymbol{\phi}_{1}\left|\mathcal{M}_{3}\right.\right)\mathcal{P}\left(\boldsymbol{\phi}_{2}\left|\mathcal{M}_{3}\right.\right)d\boldsymbol{\phi}_{1}d\boldsymbol{\phi}_{2}
\]

The un-normalised posterior distribution of $\alpha$ is given by 

\[
\mathcal{P}\left(\alpha\left|\mathcal{D},\,\mathcal{M}_{3}\right.\right)=k\,\sqrt{\left|2\pi\mathbf{P}\right|\left|2\pi\mathbf{Q}\right|}\,\textrm{exp}\left[\dfrac{1}{2}\left\{ \alpha^{2}\mathbf{A}+\left(\alpha^{2}\mathbf{B}-\mathbf{D}_{1}^{\textrm{T}}\mathbf{b}\right)^{\textrm{T}}\mathbf{Q}\left(\alpha^{2}\mathbf{B}-\mathbf{D}_{1}^{\textrm{T}}\mathbf{b}\right)\right\} \right]
\]

where 

\[
\mathbf{A}=\mathbf{b}^{\textrm{T}}\mathbf{D}_{2}\mathbf{P}\mathbf{D}_{2}^{\textrm{T}}\mathbf{b}
\]

\[
\mathbf{B}=\mathbf{D}_{1}^{\textrm{T}}\mathbf{D}_{2}\mathbf{P}\mathbf{D}_{2}^{\textrm{T}}\mathbf{b}
\]

\[
\mathbf{P}=\left(\alpha^{2}\mathbf{D}_{2}^{\textrm{T}}\mathbf{D}_{2}+\mathbf{C}_{2}^{-1}\right)^{-1}
\]

\[
\mathbf{Q}=\left(\mathbf{D}_{1}^{\textrm{T}}\mathbf{D}_{1}-\alpha^{2}\mathbf{D}_{1}^{\textrm{T}}\mathbf{D}_{2}\mathbf{P}\mathbf{D}_{2}^{\textrm{T}}\mathbf{D}_{1}+\mathbf{C}_{1}^{-1}\right)^{-1}
\]

The normalisation constant $k$ is found using Simpson's rule as it
is difficult to obtain it analytically.

%\section*{Appendix D - Combined Likelihood Method}
\section{Combined Likelihood - linear model}
\label{combined_likelihood_derivation}

The combined likelihood is given by 

\[
\mathcal{L}_{3}=\alpha\mathcal{L}_{1}+\left(1-\alpha\right)\mathcal{L}_{2}
\]

and the posterior distribution of $\alpha$

\[
\mathcal{P}\left(\alpha\left|\mathcal{D},\mathcal{M}_{1},\mathcal{M}_{2}\right.\right)=c\int_{\boldsymbol{\phi}_{1}}\int_{\boldsymbol{\phi}_{2}}\left[\alpha\mathcal{L}_{1}+\left(1-\alpha\right)\mathcal{L}_{2}\right]\mathcal{P}\left(\alpha\left|\mathcal{M}_{1},\mathcal{M}_{2}\right.\right)\mathcal{P}\left(\boldsymbol{\phi}_{1}\left|\mathcal{M}_{1},\mathcal{M}_{2}\right.\right)\mathcal{P}\left(\boldsymbol{\phi}_{2}\left|\mathcal{M}_{1},\mathcal{M}_{2}\right.\right)d\boldsymbol{\phi}_{1}d\boldsymbol{\phi}_{2}
\]

where 

\[
\mathcal{L}_{1}\sim\textrm{exp}\left[-\dfrac{1}{2}\left(\mathbf{b}-\mathbf{D}_{1}\boldsymbol{\phi}_{1}-\mathbf{D}_{2}\boldsymbol{\phi}_{2}\right)^{\textrm{T}}\left(\mathbf{b}-\mathbf{D}_{1}\boldsymbol{\phi}_{1}-\mathbf{D}_{2}\boldsymbol{\phi}_{2}\right)\right]
\]

\[
\mathcal{L}_{2}\sim\textrm{exp}\left[-\dfrac{1}{2}\left(\mathbf{b}-\mathbf{D}_{1}\boldsymbol{\phi}_{1}\right)^{\textrm{T}}\left(\mathbf{b}-\mathbf{D}_{1}\boldsymbol{\phi}_{1}\right)\right]
\]

and where $\mathbf{D}_{1}$ and $\mathbf{D}_{2}$
are the appropriate design matrices, as before and $c$ is simply is normalisation constant. We can further express
$\mathcal{L}_{1}$ in term of $\mathcal{L}_{2}$ as 

\[
\mathcal{L}_{1}\sim\mathcal{L}_{2}\textrm{exp}\left[-\dfrac{1}{2}\left(\boldsymbol{\phi}_{2}\mathbf{D}_{2}^{\textrm{T}}\mathbf{D}_{2}\boldsymbol{\phi}_{2}-2\boldsymbol{\phi}_{2}^{\textrm{T}}\mathbf{D}_{2}^{\textrm{T}}\left(\mathbf{b}-\mathbf{D}_{1}\boldsymbol{\phi}_{1}\right)\right)\right]
\]

Then 

\[
\mathcal{L}_{3}=\mathcal{L}_{2}\left\{ \alpha\,\textrm{exp}\left[-\dfrac{1}{2}\left(\boldsymbol{\phi}_{2}\mathbf{D}_{2}^{\textrm{T}}\mathbf{D}_{2}\boldsymbol{\phi}_{2}-2\boldsymbol{\phi}_{2}^{\textrm{T}}\mathbf{D}_{2}^{\textrm{T}}\left(\mathbf{b}-\mathbf{D}_{1}\boldsymbol{\phi}_{1}\right)\right)\right]+1-\alpha\right\} 
\]

The normalised posterior distribution of $\alpha$ is given by 

\[
\mathcal{P}\left(\alpha\left|\mathcal{D},\,\mathcal{M}_{1},\,\mathcal{M}_{2}\right.\right)=\dfrac{2\left(P-Q\right)}{P+Q}\alpha+\dfrac{2Q}{P+Q}
\]

where 

\[
P=\sqrt{\left|2\pi\Sigma_{1}^{-1}\right|\left|2\pi\Sigma_{2}^{-1}\right|}\,\textrm{exp}\left[-\dfrac{1}{2}\left(\mathbf{k}_{2}+\mathbf{b}^{\textrm{T}}\mathbf{b}-\mathbf{b}^{\textrm{T}}\mathbf{D}_{2}\Sigma_{1}^{-1}\mathbf{D}_{2}^{\textrm{T}}\mathbf{b}\right)\right]
\]

\[
{\displaystyle Q}=\sqrt{\left|2\pi\Sigma_{3}^{-1}\right|\left|2\pi\mathbf{C}_{2}^{-1}\right|}\,\textrm{exp}\left[-\dfrac{1}{2}\left(\mathbf{k}_{3}+\mathbf{b}^{\textrm{T}}\mathbf{b}\right)\right]
\]

and 

\[
{\displaystyle \Sigma}_{1}=\mathbf{v}^{\textrm{T}}\mathbf{v}+\mathbf{C}_{2}^{-1}
\]

\[
\Sigma_{2}=\mathbf{C}_{1}^{-1}-\mathbf{D}_{1}^{\textrm{T}}\mathbf{D}_{2}\Sigma_{1}^{-1}\mathbf{D}_{2}^{\textrm{T}}\mathbf{D}_{1}+\mathbf{D}_{1}^{\textrm{T}}\mathbf{D}_{1}
\]

\[
\Sigma_{3}=\mathbf{D}_{1}^{\textrm{T}}\mathbf{D}_{1}+\mathbf{C}_{1}^{-1}
\]

\[
\mathbf{k}_{2}=-\left(\mathbf{D}_{1}^{\textrm{T}}\mathbf{D}_{2}\Sigma_{1}^{-1}\mathbf{D}_{2}^{\textrm{T}}\mathbf{b}-\mathbf{D}_{1}^{\textrm{T}}\mathbf{b}\right)^{\textrm{T}}\Sigma_{2}^{-1}\left(\mathbf{D}_{1}^{\textrm{T}}\mathbf{D}_{2}\Sigma_{1}^{-1}\mathbf{D}_{2}^{\textrm{T}}\mathbf{b}-\mathbf{D}_{1}^{\textrm{T}}\mathbf{b}\right)
\]

\[
\mathbf{k}_{3}=-\left(\mathbf{D}_{1}^{\textrm{T}}\mathbf{b}\right)^{\textrm{T}}\Sigma_{3}^{-1}\left(\mathbf{D}_{1}^{\textrm{T}}\mathbf{b}\right)
\]

Hence, the Bayes Factor is given by 

\[
B_{21}=\dfrac{\mathcal{P}\left(\alpha=0\left|\mathcal{D},\,\mathcal{M}_{1},\,\mathcal{M}_{2}\right.\right)}{\mathcal{P}\left(\alpha=1\left|\mathcal{D},\,\mathcal{M}_{1},\,\mathcal{M}_{2}\right.\right)}=\dfrac{Q}{P}
\]

\section{Bayes factor precision in the combined likelihood approach}
\label{precision_bayes_factor}

The posterior distribution of $\alpha$ can be written as $\left(2-2c\right)\alpha+c$
and the log-Bayes Factor as $\textrm{log B}_{21}=\textrm{log}\left(\dfrac{c}{2-c}\right)$.
The error in $\textrm{log B}_{21}$ with respect to $c$ is 

\[
\sigma_{\textrm{log B}_{21}}^{2}=\left[\dfrac{2}{c\left(2-c\right)}\right]^{2}\sigma_{c}^{2}
\]

Moreover, if we assume that the error in each bin can be modelled
using Poisson statistics, it can be shown that 

\[
\sigma_{\textrm{log B}_{21}}^{2}=\dfrac{4m}{Nc^{2}\left(2-c\right)^{2}}\sum_{i=1}^{m}\dfrac{n_{i}}{\left(1-2\alpha_{i}\right)^{2}}
\]

where $m$ is the number of bins and $N$ is the total number of samples.
Hence, 

\[
\sigma_{\textrm{log B}_{21}}\propto\dfrac{1}{\sqrt{N}c\left(2-c\right)}
\]

\end{document}